\documentclass[prd,twocolumn,superscriptaddress,altaffilletter,nofootinbib]{revtex4}

\usepackage{cancel}
\usepackage{graphicx}
\usepackage{amsmath}
\usepackage{amssymb}
\usepackage{graphicx,epsfig}
\newcommand{\be}{\begin{equation}}
\newcommand{\ee}{\end{equation}}
\newcommand{\bea}{\begin{eqnarray}}
\newcommand{\eea}{\end{eqnarray}}
\newcommand{\der}{\partial}
\newcommand{\vphi}{\varphi}

\begin{document}

%%%%%%%%%%%%%%%%%%%%%%%%%%

%%%%%%%%%%%%%%%%%%%%%%%%%%%%%%%%%%%%%%%%%%%%%%%%%%%%%%%%%%%%%%%%%%%%%%%%%%%%%%

\title{On the role of postulates in the understanding of Weyl gauge symmetry}

%%%%%%%%%%%%%%%%%%%%%%%%%%%%%%%%%%%%%%%%%%%%%%%%%%%%%%%%%%%%%%%%%%%%%%%%%%%%%%

\author{Israel Quiros}\email{iquiros@fisica.ugto.mx}\affiliation{Dpto. Ingenier\'ia Civil, Divisi\'on de Ingenier\'ia, Universidad de Guanajuato, Campus Guanajuato, Gto., M\'exico.}

\date{\today}

\begin{abstract} Understanding of Weyl gauge symmetry is rarely associated with underlying postulates. Here we show that such an omission leads to discrepancies in regard to the reach and consequences of gauge symmetry within gravitational theories. Replacement of the postulates which underlie the conventional approach, by just the opposite assumptions, leads to a radically different interpretation of Weyl gauge symmetry, which shares the same mathematical foundations and carries observational consequences.\end{abstract}

%\pacs{04.50.Kd, 04.50.Cd, 11.10.Ef, 98.80.-k, 98.80.Jk}

%%%%%%%%%%%%%%%%%%%%%%%%%%%%%%%%%%%%%%%%

\maketitle

%%%%%%%%%%%%%%%%%%%%%%%%%%%%%%%%%%%%%%%

%--------------Beginning of draft--------------

%%%%%%%%%%%%%%%%%%%%%%%%%%%%%%%%%%%%%%%%%%%%%%%%%

\section{Introduction}\label{sect-intro}

%%%%%%%%%%%%%%%%%%%%%%%%%%%%%%%%%%%%%%%%%%%%%%%%%

Weyl geometry \cite{weyl-1918, weyl-book, london-1927, schouten-book, dirac-1973, utiyama-1973, adler-book, maeder-1978, smolin-1979, cheng, tamm-1999, many-weyl-book} and its associated Weyl gauge symmetry (WGS), gave rise to the advent of the gauge theories of the fundamental interactions. However, it seems that classical gravitational theories, where the gauge history began, do not require of WGS, apart from, perhaps, short range and early time gravitational phenomena \cite{smolin-1979, cheng, tamm-1999, many-weyl-book, thooft-2015, ghilen-prd, ghilen-1, ghilen-2}. It is widely accepted that WGS must be, at most, a badly broken symmetry in Nature. However this viewpoint is not accepted by everyone. As stated in \cite{comment-2} (see also \cite{shaukat-1, shaukat-2, shaukat-3}): ``Weyl invariance seems not to be a manifest symmetry of nature. But neither are $U(1)$ or general coordinate invariance..., if one picks a particular gauge. Perhaps physical theories have been written in a special Weyl gauge thus breaking the Weyl symmetry!''

The relevance of Weyl gauge transformations (WGT), also known as local scale transformations and associated with the transformations of units, was anticipated in Ref. \cite{dicke-1962}, where it was proposed the following postulate: ``...the particular values of the units of mass, length, and time employed are arbitrary...the laws of physics must be invariant under a transformation of units.'' It is also recognized, for instance, in the essay \cite{thooft-2015}, where it is conjectured that: ``Small time and distance scales seem not to be related to large time and distance scales...because we fail to understand the symmetry of the scale transformations.'' 

%------------------------------------------------------

Frequently, disagreements in the understanding of given approach come from ignorance of the postulates on which the mathematical structure of the approach is based. This is, perhaps, the origin of the longstanding discussion in the bibliography on the role of the conformal frames (among them the Jordan and Einstein frames) within scalar-tensor theories (STT) \cite{dicke-1962, faraoni-2007, fujii-book, quiros-rev, faraoni_rev_1997, faraoni_ijtp_1999, ct-ineq-nojiri, ct-ineq-brisc, sarkar_mpla_2007, ct-ineq-capoz, quiros-grg-2013, fatibene-2014, ct(inequiv)-7, ct-ineq-brooker, ct-ineq-baha, ct(inequiv)-8, paliatha-2, quiros-ijmpd-2020, ct-ineq-2020, faraoni-book, flanagan-2004, sotiriou_etall_ijmpd_2008, saal_cqg_2016, ct-1, ct(fresh-view)-3, indios_consrvd_prd_2018, thermod_prd_2018, shtanov-2022}. Many researchers agree that the Einstein and Jordan frames are not physically equivalent \cite{faraoni_rev_1997, faraoni_ijtp_1999, ct-ineq-nojiri, ct-ineq-brisc, sarkar_mpla_2007, ct-ineq-capoz, quiros-grg-2013, fatibene-2014, ct(inequiv)-7, ct-ineq-brooker, ct-ineq-baha, ct(inequiv)-8, paliatha-2, quiros-ijmpd-2020, ct-ineq-2020}, while many others support quite the contrary standpoint: that these are physically equivalent frames \cite{faraoni-book, flanagan-2004, sotiriou_etall_ijmpd_2008, saal_cqg_2016, ct-1, ct(fresh-view)-3, indios_consrvd_prd_2018, thermod_prd_2018, shtanov-2022}. Possibly both points of view are correct, just different starting premises are assumed. 

For the above reason we think that, investigating the role of postulates in the understanding of WGS, is of primordial importance for the correct assessment of the reach and possible consequences of this symmetry. In the present paper we investigate the geometrical and physical consequences of two opposite approaches to WGS: the conventional one and an alternative approach.

%--------------------------------------------------------------

The conventional approach to Weyl gauge symmetry is based on the following postulates: 1) WGS is a redundancy of the mathematical description of gravitational phenomena, so that only Weyl gauge invariant (WGI) quantities have physical meaning, and 2) dimensionful constants are not allowed, since these break WGS \cite{smolin-1979, cheng, tamm-1999, bars, bars-sailing, thooft-2015, ghilen-prd, ghilen-1, ghilen-2}. As consequence of the above postulates it follows, for instance, that Weyl integrable geometry (WIG)\footnote{Do not confound WGI (for Weyl gauge invariant or Weyl gauge invariance, etc.) with WIG (for Weyl integrable geometry).} is trivial because it does not carry independent geometrical meaning beyond that of Riemann geometry \cite{schouten-book, adler-book}. In what follows, for shorthand, we shall call this approach as the ``CAP,'' (for ``conventional approach.'') Although the mathematical basis of the CAP is solid, this approach is based on postulates, which are not unique.  

Despite of the CAP, there are several works in the bibliography where it is assumed that WIG has independent geometrical meaning \cite{wig1, wig2, wig3, wig4, wig5, wig6, wig7, wig8, wig9, wig10, wig11, wig12, wig13, wig14, wig15, wig16, wig17, wig18, wig19, wig20}. The mathematical basement for these works is less known. As a matter of fact, in most of the works where WIG space, denoted here by $\tilde W_4$, is the basis for physical explanation of given phenomena, neither the associated WGS nor the mathematical foundations for the choice of $\tilde W_4$, are discussed.

Our motivation for the present paper is that, since the postulates that underlie the CAP's theoretical framework are not unique, these may well be replaced by just the opposite postulates, thus leading to a radically different approach which meets the same mathematical foundations. In this paper the CAP's postulates are replaced by the following assumptions: 1) WGS is not a redundancy of the mathematical description of the gravitational laws, and 2) neither the masses of fields, nor the dimensionful constants, break WGS. In what follows, for shorthand, we call the resulting framework as the ``AAP'' (for ``alternative approach.'') In the present work our emphasis is in the demonstration of the non-triviality of WGS and in the search for its physical consequences, through following the AAP. 

%---------------------------------organization----------------------------

The outline of the paper is the following. The basic elements of WGS and of the theory of gravity subject of the present research, are exposed in section \ref{sect-basics}, while in section \ref{sect-convent-app} we discuss on the impact of the underlying postulates in the geometrical and physical consequences of the CAP. It will be shown, in particular, that the assumption that WGS is a mathematical redundancy, leads to coarse trivialization of the Weyl gauge transformations. In section \ref{sect-new-app} the postulates that underlie the AAP are discussed while in section \ref{sect-gtheor} the consequences of the AAP for the gravitational theories are investigated. In section \ref{sect-z} we explore the possibility that, thanks to the redshift effect, the AAP might have experimental support. In section \ref{sect-critique} we address several critical comments on the present approach to WGS, that either have arisen or might arise during scientific discussion on the subject. The results of this paper are discussed in section \ref{sect-discu} while brief conclusions are given in section \ref{sect-conclu}. 

In order to complement the mathematical foundations of the paper an appendix section has been included. A brief demonstration that varying masses in $V_4$ space lead to a fifth force is given in appendix \ref{app-a}. In appendix \ref{app-b} the mathematical derivation of the time-like autoparallel's equation in Weyl space $W_4$, is given. The particular case of $\tilde W_4$ space is also discussed. Meanwhile, in appendix \ref{app-c}, the derivation of the expression for the point-dependent mass is exposed. This latter expression is important for the understanding of the discussion about the shift of frequency.

%%%%%%%%%%%%%%%%%%%%%%%

\section{The basics}
\label{sect-basics}

%%%%%%%%%%%%%%%%%%%%%%

The nonmetricity law of $W_4$ space \cite{weyl-1918, weyl-book, london-1927, schouten-book, dirac-1973, utiyama-1973, adler-book, maeder-1978, smolin-1979, cheng, tamm-1999, many-weyl-book}, reads: $\hat\nabla_\alpha g_{\mu\nu}=-Q_\alpha g_{\mu\nu},$ where $Q_\mu$ is the nonmetricity vector and the covariant derivative $\hat\nabla_\mu$ is defined with respect to the torsion-free affine connection of the manifold: 

\begin{align} \Gamma^\alpha_{\;\;\mu\nu}=\{^\alpha_{\mu\nu}\}+L^\alpha_{\;\;\mu\nu},\label{aff-c}\end{align} where 

\begin{align} \{^\alpha_{\mu\nu}\}:=\frac{1}{2}g^{\alpha\lambda}\left(\der_\nu g_{\mu\lambda}+\der_\mu g_{\nu\lambda}-\der_\lambda g_{\mu\nu}\right),\label{lc-c}\end{align} is the Levi-Civita (LC) connection, while 

\begin{align} L^\alpha_{\;\;\mu\nu}:=\frac{1}{2}\left(Q_\mu\delta^\alpha_\nu+Q_\nu\delta^\alpha_\mu-Q^\alpha g_{\mu\nu}\right),\nonumber\end{align} is the disformation tensor. The Weyl gauge vector $Q_\mu$ measures how much the length of given vector varies during parallel transport. When the nonmetricity vector amounts to the gradient of a scalar: $Q_\mu=\der_\mu\phi^2/\phi^2=2\der_\mu\phi/\phi,$ the vectorial nonmetricity law is replaced by

\bea \hat\nabla_\alpha g_{\mu\nu}=-\frac{\der_\alpha\phi^2}{\phi^2}g_{\mu\nu}=-2\frac{\der_\alpha\phi}{\phi}g_{\mu\nu}.\label{grad-nm}\eea The spaces where the gradient nonmetricity condition \eqref{grad-nm} is satisfied, are the $\tilde W_4$ spaces \cite{wig1, wig2, wig3, wig4, wig5, wig6, wig7, wig8, wig9, wig10, wig11, wig12, wig13, wig14, wig15, wig16, wig17, wig18, wig19, wig20}. In this case the covariant derivative $\hat\nabla_\alpha$ is defined with respect to the affine connection \eqref{aff-c} where the disformation tensor is given by:

\begin{align} L^\alpha_{\;\;\mu\nu}=\phi^{-1}\left(\der_\mu\phi\delta^\alpha_\nu+\der_\nu\phi\delta^\alpha_\mu-\der^\alpha\phi g_{\mu\nu}\right).\label{disf}\end{align} 

Weyl gauge invariance is one of the most important properties of $\tilde W_4$ spaces, which are a subset of $W_4$. This symmetry arises from invariance of the nonmetricity law \eqref{grad-nm}, of the affine connection $\Gamma^\alpha_{\;\;\mu\nu}$ and of the related curvature tensor $\hat R^\alpha_{\;\;\mu\beta\nu}$, etc., under the WGT:\footnote{Weyl gauge transformations \eqref{gauge-t} are not diffeomorphisms. The spacetime coordinates are not modified by the WGT because these act only on fields.}

\bea g_{\mu\nu}\rightarrow\Omega^2g_{\mu\nu},\;\phi\rightarrow\Omega^{-1}\phi,\;\Psi_A\rightarrow\Omega^{w_A}\Psi_A,\label{gauge-t}\eea where the $\Psi_A$ represent non gravitational fields with conformal weights $w_A$, living in the background space $\tilde W_4$. 

%----------------------Lagrangian------------------------------------

In the present paper we adhere to the premise stated in \cite{shaukat-1, shaukat-2, shaukat-3}, that ``physical theories have been written in a special Weyl gauge'' and that ``theories with dimensionful couplings are the gauge fixed versions of certain Weyl invariant ones,'' although we do not follow here the route taken by the authors of \cite{shaukat-1, shaukat-2, shaukat-3}, where the tractor calculus is undertaken in order to look for physical consequences of Weyl gauge invariance, in particular, for fermionic and interacting supersymmetric theories. 

In this paper we take a new look at well-known WGI theory of gravity, which is given by the gravitational Lagrangian \cite{ghilen-prd, ghilen-1, ghilen-2, bars, bars-sailing, deser-1970, quiros-2014, quiros-prd-2023-1, quiros-arxiv, jackiw-2014, oda-2022, rodrigo-arxiv}:

\begin{align} L_\text{grav}=\frac{\sqrt{-g}}{2}\left[\phi^2 R+6(\der\phi)^2-\frac{\lambda}{4}\phi^4\right],\label{grav-lag}\end{align} where $R$ is the curvature scalar of Riemann space $V_4$, $\lambda$ is a dimensionless constant and we use the notation $(\der\phi)^2\equiv g^{\mu\nu}\der_\mu\phi\der_\nu\phi$. Here we investigate this well-known Lagrangian in connection with $V_4$ and also with $\tilde W_4$ spaces.

%%%%%%%%%%%%%%%%%%%%%%%%%%%%%%%%%%%%%%%

\section{Conventional approach}
\label{sect-convent-app}

%%%%%%%%%%%%%%%%%%%%%%%%%%%%%%%%%%%%%%

%---------------------First postulate-----------------

According to the first CAP's postulate the WGS must be interpreted as a mathematical redundancy. Means that the different gauges may be identified, so that the gauge choice has no physical consequences: The quantities with the physical meaning are necessarily WGI quantities. In particular, the physically meaningful metric is the gauge invariant product: $\mathfrak{g}_{\mu\nu}\equiv\phi^2g_{\mu\nu}$. Hence, neither the starting metric $g_{\mu\nu}$, nor the scalar field $\phi$, have independent physical meaning. For that reason two apparently different states: 

\begin{align} {\cal G}_a=\left({\cal M}_4,g^{(a)}_{\mu\nu},\phi_a\right)=\left({\cal M}_4,\mathfrak{g}^{(a)}_{\mu\nu}\right),\nonumber\end{align} and 

\begin{align} {\cal G}_b=\left({\cal M}_4,g^{(b)}_{\mu\nu},\phi_b\right)=\left({\cal M}_4,\mathfrak{g}^{(b)}_{\mu\nu}\right),\nonumber\end{align} must be identified: ${\cal G}_a\equiv{\cal G}_b$, as long as the product $\phi^2g_{\mu\nu}$, is WGI:

\begin{align} \phi_a^2g^{(a)}_{\mu\nu}=\phi_b^2g^{(b)}_{\mu\nu}\;\Rightarrow\;\mathfrak{g}^{(a)}_{\mu\nu}=\mathfrak{g}^{(b)}_{\mu\nu}.\nonumber\end{align} In consequence the physically meaningful line element is:

\begin{align} d\mathfrak{s}^2=\mathfrak{g}_{\mu\nu}dx^\mu dx^\nu,\label{triv-le}\end{align} while the measured element of proper time: $\Delta\mathfrak{t}=i\int d\mathfrak{s}.$ The WGI curvature invariants, such as the curvature scalar and the the Kretschmann invariant of the physical metric are given by the following expressions,

\begin{align}\mathfrak{R}=\mathfrak{g}^{\mu\nu}\mathfrak{R}_{\mu\nu},\;K(\mathfrak{g}):=\mathfrak{R}^{\sigma\mu\lambda\nu}\mathfrak{R}_{\sigma\mu\lambda\nu},\nonumber\end{align} where $\mathfrak{R}_{\mu\nu}=\mathfrak{g}^{\lambda\sigma}\mathfrak{R}_{\lambda\mu\sigma\nu}$ is the Ricci tensor and $\mathfrak{R}^\alpha_{\;\;\mu\beta\nu}$ is the Riemann-Christoffel curvature tensor of the physical metric $\mathfrak{g}_{\mu\nu}.$ These quantities are defined with respect to the affine connection: 

\begin{align}\mathfrak{C}^\alpha_{\;\;\mu\nu}:=\frac{1}{2}\mathfrak{g}^{\alpha\lambda}\left(\der_\nu\mathfrak{g}_{\mu\lambda}+\der_\mu\mathfrak{g}_{\nu\lambda}-\der_\lambda\mathfrak{g}_{\mu\nu}\right),\label{triv-lc-c}\end{align} which coincides with the Levi-Civita connection of the physical metric. 

According to the CAP, the Weyl gauge transformations \eqref{gauge-t}, collapse to the trivial identity transformation: 

\begin{align} &\mathfrak{g}_{\mu\nu}\rightarrow\mathfrak{g}_{\mu\nu},\;\sqrt{-\mathfrak{g}}\rightarrow\sqrt{-\mathfrak{g}},\;\mathfrak{C}^\alpha_{\;\;\mu\nu}\rightarrow\mathfrak{C}^\alpha_{\;\;\mu\nu},\nonumber\\ &\mathfrak{R}^\alpha_{\;\;\mu\beta\nu}\rightarrow\mathfrak{R}^\alpha_{\;\;\mu\beta\nu},\;\mathfrak{R}_{\mu\nu}\rightarrow\mathfrak{R}_{\mu\nu},\;\mathfrak{R}\rightarrow\mathfrak{R}.\label{triv-g-t}\end{align} This is consequence of the first postulate which implicitly amounts to the metric $g_{\mu\nu}$ and the compensator field $\phi$ in the gravitational Lagrangian \eqref{grav-lag}, being not independent fields, since the one with the physical meaning is their product $\mathfrak{g}_{\mu\nu}.$ 

Let us take, for instance, the Einstein-Hilbert (EH) Lagrangian for the physically meaningful metric: 

\begin{align} L_\text{eh}=\frac{1}{2}\sqrt{-\mathfrak{g}}\mathcal{R},\label{triv-eh-lag}\end{align} over $V_4$ spacetime, (${\cal M}_4,\mathfrak{g}_{\mu\nu}$.) This means that the metricity condition: $\mathcal{D}_\alpha\mathfrak{g}_{\mu\nu}=0$, is satisfied, where $\mathcal{D}_\mu$ is the covariant derivative defined with respect to the LC connection of the physical metric \eqref{triv-lc-c}. Notice that the explicit effect of the compensator field, as well as that of the starting metric, have been removed from the physical description. We end up considering just general relativity (GR) theory over $V_4$ spacecime (${\cal M}_4,\mathfrak{g}_{\mu\nu}$,) where ${\cal M}_4\in V_4$ with the WGT \eqref{gauge-t} collapsed to the unity transformation \eqref{triv-g-t}. This amounts to coarse trivialization of the WGT and their role in gravitation.

%-----------------------------second postulate-----------------------

In addition to the requirement that WGS must be a mathematical redundancy, the CAP assumes that dimensionful constants are not allowed, since these break WGS \cite{smolin-1979, cheng, tamm-1999, bars, bars-sailing, ghilen-prd, ghilen-1, ghilen-2}. In particular a mass parameter would break gauge invariance. A typical example is given in Ref. \cite{deser-1970}, where in order to break the WGI of the gravitational Lagrangian \eqref{grav-lag}, the following symmetry breaking term is added:

\begin{align} L_{\phi^2}=-\frac{\sqrt{-g}}{2}m^2_\phi\phi^2.\label{sb-t}\end{align} Here the mass of the scalar field $m_\phi$ is assumed to be unchanged by the gauge transformations \eqref{gauge-t}, which is a reasonable assumption in $V_4$ space, where the masses of point particles are constant over spacetime.

%---------------------trivial WIG---------------------

According to the CAP, since $Q_\mu=\der_\mu\phi^2/\phi^2$ is not dynamical, under the choice $\phi=$const., the $\tilde W_4$ space becomes into $V_4$ \cite{schouten-book, adler-book, ghilen-2}. The implicit conclusion is that Weyl integrable geometry does not carry independent geometrical meaning beyond that of Riemann geometry, i. e., WIG is trivial: it is ``pure gauge.'' Let us expose the analysis that leads to this conclusion. It has been stated in \cite{adler-book} that, due to the Stokes theorem, vanishing of the quantity $Q_{\mu\nu}:=2\der_{[\mu}Q_{\nu]}=0$, is the necessary and sufficient condition for the given vector to return to itself after parallel transport in a closed trajectory in $W_4$ space:\footnote{The use of the Stokes theorem \eqref{oint}, in a real physical situation would require the given body (for instance, an atomic clock) to be submitted to parallel transport in a closed timelike trajectory ${\cal C}$ in spacetime. In general, closed timelike curves (CTCs) in spacetime carry causality issues as long as a CTC represents time travel \cite{ctc-morris, ctc-friedman, ctc-thorne, ctc-cho, ctc-bonnor, ctc-luminet}.}

\bea \oint_{\cal C}Q_\mu dx^\mu=\frac{1}{2}\iint_SQ_{\mu\nu}dx^\mu\wedge dx^\nu,\label{oint}\eea where $\cal C$ is the boundary of the oriented surface $S$. If $Q_{\mu\nu}=0$, then $\oint Q_\mu dx^\mu=0$. In this case the length of the vector upon completing the closed path: $l=l_0\exp\oint_{\cal C}Q_\mu dx^\mu=l_0$, is unchanged. Alternatively, it is said that the condition $\der_{[\mu}Q_{\nu]}=0$, entails that WIG is reducible to Riemann geometry. An additional geometrical implication of the non-metricity law \eqref{grad-nm} is that the length of vectors is point-dependent, as for instance, in $\tilde W_4$ space, where $Q_\mu=\der_\mu\ln\phi^2$: 

\begin{align} l(x)=l_0\exp\int_0^x\der_\mu(\ln\phi^2)dx^\mu=l_0\phi^2(x)/\phi^2(0),\label{length-point-dep}\end{align} where $l_0$ is the length of the vector at the origin. In this case the requirement $\der_{[\mu}Q_{\nu]}=0$, which is obviously satisfied in $\tilde W_4$ space, does not warrant that the resulting space is trivially $V_4$ spacetime. 

Only because of the CAP assumption that WGS is a mathematical redundancy, which means that the gauge choice loses any physical significance, one may safely fix the gauge scalar to be a constant in spacetime: $\phi(x)=\phi_0=$const., so that from \eqref{length-point-dep} it follows that $l(x)=l_0$, meaning that Riemann geometry is recovered. 

Summarizing: WIG is trivial and can not be differentiated from $V_4$ if the condition $\der_{[\mu}Q_{\nu]}=0$, takes place and, simultaneously, WGS is assumed as a mathematical redundancy.

%==============================================================

\subsection{Implications of the CAP for gravitational theory}

%==============================================================

The investigation of the physical consequences of WGS requires of specific theories. Accordingly, here we study the consequences of the CAP in the ``linear'' WGI gravitational theory, which is given by the gravitational Lagrangian \eqref{grav-lag} \cite{ghilen-prd, ghilen-1, ghilen-2, bars, bars-sailing, deser-1970, quiros-prd-2023-1, quiros-arxiv, jackiw-2014, oda-2022, rodrigo-arxiv}. We shall consider two different scenarios: 1) when the Lagrangian \eqref{grav-lag} operates in $V_4$ background space, as in \cite{bars, bars-sailing, deser-1970}, and 2) when the above gravitational Lagrangian is associated with $\tilde W_4$ background space \cite{ghilen-prd, ghilen-1, ghilen-2, quiros-prd-2023-1, quiros-arxiv}. 

The LC connection in terms of the WGI physical metric $\mathfrak{g}_{\mu\nu}=\phi^2g_{\mu\nu}$, is given by Eq. \eqref{triv-lc-c}. It coincides with the affine connection of $\tilde W_4$ space: $\mathfrak{C}^\alpha_{\;\;\mu\nu}=\Gamma^\alpha_{\;\;\mu\nu},$ where

\begin{align} \Gamma^\alpha_{\;\;\mu\nu}=\{^\alpha_{\mu\nu}\}+\frac{1}{\phi}\left(\der_\mu\phi\delta^\alpha_\nu+\der_\nu\phi\delta^\alpha_\mu-\der^\alpha\phi g_{\mu\nu}\right).\label{wig-aff-c}\end{align} This means, in turn, that the Riemann-Christoffel curvature tensor of the physical metric coincides with the curvature tensor of $\tilde W_4$ space: $\mathfrak{R}^\alpha_{\;\;\mu\sigma\nu}=\hat R^\alpha_{\;\;\mu\sigma\nu}$, while $\mathfrak{R}_{\mu\nu}=\hat R_{\mu\nu}$. In consequence:

\begin{align} \mathfrak{R}=\phi^{-2}\hat R=\phi^{-2}\left[R-3\frac{(\der\phi)^2}{\phi^2}-6\nabla_\mu\left(\frac{\der^\mu\phi}{\phi}\right)\right],\label{wig-eff}\end{align} where the quantities with the hat account for $\tilde W_4$ quantities and operators, while the ones without the hat amount to Riemannian objects. Hence, up to total derivatives: $\sqrt{-\mathfrak{g}}\mathfrak{R}=\sqrt{-g}\left[\phi^2R+6(\der\phi)^2\right].$ Means that, in terms of the physical metric $\mathfrak{g}_{\mu\nu}$, the gravitational Lagrangian \eqref{grav-lag} reads:

\begin{align} L_\text{grav}=\frac{\sqrt{-\mathfrak{g}}}{2}\left[\mathfrak{R}-2\Lambda\right],\label{triv-grav-lag}\end{align} where $\Lambda=\lambda/8$.

%----------------------------------------

\subsubsection{$V_4$ background space}

%----------------------------------------

When the Lagrangian \eqref{grav-lag}, is associated with $V_4$ background space, this means that timelike point-particles follow geodesics of $V_4$ space:

\begin{align} \frac{d^2x^\alpha}{ds^2}+\left\{^\alpha_{\mu\nu}\right\}\frac{dx^\mu}{ds}\frac{dx^\nu}{ds}=0.\label{t-l-geod}\end{align} This equation is not invariant under the WGT \eqref{gauge-t}. This is manifest if write Eq. \eqref{t-l-geod} in terms of the WGI metric $\mathfrak{g}_{\mu\nu},$ which is the physically meaningful one. Eq. \eqref{t-l-geod} can then be written as:

\begin{align} \frac{d^2x^\alpha}{d\mathfrak{s}^2}+\mathfrak{C}^\alpha_{\mu\nu}\frac{dx^\mu}{d\mathfrak{s}}\frac{dx^\nu}{d\mathfrak{s}}+\mathfrak{h}^{\alpha\mu}\frac{\der_\mu\phi}{\phi}=0,\label{triv-tl-geod}\end{align} where $\mathfrak{h}^{\mu\alpha}=\phi^{-2}h^{\mu\alpha}$ and the orthogonal projection tensor $h^{\mu\alpha}$ is defined in \eqref{orto-proj}. The last term in the LHS of the above equation is the one which spoils the WGS. In this case the timelike point-particles (and fields) do not follow geodesics of Riemann space $({\cal M}_4,\mathfrak{g}_{\mu\nu}).$ The fifth force $\propto\der^\alpha\phi/\phi$, deviates these particles from geodesic motion (see the related discussion in appendix \ref{app-a}.) This force could be measured in terrestrial and Solar system experiments \cite{will-lrr}.

The above difficulty is circumvented thanks to a well-known property of the theory \eqref{grav-lag} over $V_4$ background space \cite{quiros-prd-2023-1}: Only massless matter fields with vanishing trace of the stress-energy tensor, can couple to gravity. Means that only photons and radiation couple to gravity in this theory. As a consequence, the null geodesic equations of $V_4$ space are themselves WGI. Actually, these read, 

\bea \frac{dk^\mu}{d\xi}+\left\{^\mu_{\nu\sigma}\right\}k^\nu k^\sigma=0,\label{0-geod}\eea where $k^\mu\equiv dx^\mu/d\xi$ is the wave vector: $k_\mu k^\mu=0$, and $\xi$ is an affine parameter along null-geodesic.\footnote{Based on dimensional analysis it follows that $k^\mu$ has conformal weight $w=-2$, like the fourth-momentum $p^\mu=mdx^\mu/ds$. Hence, under \eqref{gauge-t} the wave vector transforms like $k^\mu\rightarrow\Omega^{-2}k^\mu$ $\Rightarrow$ $d\xi\rightarrow\Omega^2d\xi$, while: $dk^\mu/d\xi\rightarrow\Omega^{-4}\left[dk^\mu/d\xi-2k^\mu d\ln\Omega/d\lambda\right]$ and $\{^\alpha_{\lambda\nu}\}k^\lambda k^\nu\rightarrow\Omega^{-4}\left[\{^\mu_{\lambda\nu}\}k^\lambda k^\nu+2k^\mu d\ln\Omega/d\lambda\right],$ so that Eq. \eqref{0-geod} is not transformed by the WGT \eqref{gauge-t} (see appendix D of \cite{wald-book}.)} In terms of the physical metric $\mathfrak{g}_{\mu\nu}$, Eq. \eqref{0-geod} can be written in a manifest gauge invariant way:

\begin{align} \frac{d\mathfrak{K}^\alpha}{d\mathfrak{z}}+\mathfrak{C}^\alpha_{\mu\nu}\mathfrak{K}^\mu\mathfrak{K}^\nu=0,\label{triv-0-geod}\end{align} where we have introduced the WGI (physical) wave vector $\mathfrak{K}^\alpha=dx^\alpha/d\mathfrak{z}$ ($d\mathfrak{z}=\phi^2d\xi$ is the derivative of the WGI affine parameter along the null geodesic,) so that $\mathfrak{K}_\mu\mathfrak{K}^\mu=0.$ Notice that, under \eqref{gauge-t}: $d\mathfrak{z}\rightarrow d\mathfrak{z}$, $\mathfrak{K}^\alpha\rightarrow\mathfrak{K}^\alpha.$

Summarizing: In terms of the physically meaningful metric $\mathfrak{g}_{\mu\nu},$ the theory \eqref{grav-lag} over Riemann space $V_4$, is given by \eqref{triv-grav-lag} with background radiation obeying \eqref{triv-0-geod}. This theory is trivially gauge invariant because the Weyl rescalings \eqref{gauge-t} collapse to the identity transformations \eqref{triv-g-t}, complemented with $\mathfrak{K}^\mu\rightarrow\mathfrak{K}^\mu.$

%----------------------------------------------

\subsubsection{$\tilde W_4$ background space}

%----------------------------------------------

Let us now consider the theory \eqref{grav-lag} over $\tilde W_4$ space. In this case any matter fields, no matter whether massless or with the mass, couple to gravity. The timelike autoparallel curves of $\tilde W_4$ space satisfy (for a detailed exposition see the appendix \ref{app-b}):

\begin{align} \frac{d^2x^\alpha}{ds^2}+\left\{^\alpha_{\mu\nu}\right\}\frac{dx^\mu}{ds}\frac{dx^\nu}{ds}-\frac{\der_\mu \phi}{\phi}h^{\mu\alpha}=0,\label{time-auto-p}\end{align} where $h^{\mu\alpha}$ is the orthogonal projection tensor defined in Eq. \eqref{orto-proj}. In terms of the WGI metric $\mathfrak{g}_{\mu\nu}$ this equation can be written in a manifest WGI form:

\begin{align} \frac{d^2x^\alpha}{d\mathfrak{s}^2}+\mathfrak{C}^\alpha_{\mu\nu}\frac{dx^\mu}{d\mathfrak{s}}\frac{dx^\nu}{d\mathfrak{s}}=0,\label{triv-time-auto-p}\end{align} meanwhile, the null auto-parallels coincide with \eqref{triv-0-geod}.

Hence, in terms of the physical metric $\mathfrak{g}_{\mu\nu}$, the theory \eqref{grav-lag} over $\tilde W_4$ space, is equivalent to GR over Riemann space $V_4$, which is given by equations \eqref{triv-grav-lag}, \eqref{triv-0-geod} and \eqref{triv-time-auto-p}. In this case the WGT \eqref{gauge-t} collapse to the identity transformations \eqref{triv-g-t}, so that these play no role at all. We conclude that CAP trivializes the WGS.

%%%%%%%%%%%%%%%%%%%%%%%%%%%%%%%%%%%%%%

\section{Alternative approach}
\label{sect-new-app}

%%%%%%%%%%%%%%%%%%%%%%%%%%%%%%%%%%%%%

%-----------------------1rst potulate---------------------

Although the CAP is genuine, it is obviously not unique and it is not a fruitful approach either since, as long as the physically meaningful metric is the composite metric $\mathfrak{g}_{\mu\nu}=\phi^2g_{\mu\nu},$ the WGT amount to the trivial unity transformations \eqref{triv-g-t}: $\mathfrak{g}_{\mu\nu}\rightarrow\mathfrak{g}_{\mu\nu}$. One may assume, alternatively, that the metric $g_{\mu\nu}$ and the gauge field $\phi$ are independent fields, and that the metric has immediate physical meaning. As a consequence, not only WGI quantities are physically meaningful. Other quantities, such as the metric itself, which is transformed under \eqref{gauge-t}: $g_{\mu\nu}\rightarrow\Omega^2g_{\mu\nu}$, have manifest physical significance. This alternative assumption is not independent of the first postulate on which the AAP rests: ``WGS is not a redundancy of the mathematical description of the gravitational laws,'' since the assumption that the metric $g_{\mu\nu}$ has independent physical meaning, means that two gravitational states ``$a$'' and ``$b$,'' which are characterized by different metric tensors $g^{(a)}_{\mu\nu}$ and $g^{(b)}_{\mu\nu}$, respectively, can not be identified and so, the WGS can not be interpreted as a mathematical redundancy. 

In other words, the first AAP's postulate leads to conclude that the WGT \eqref{gauge-t}:

\begin{align} g^{(a)}_{\mu\nu}=\Omega^2_{ab}g^{(b)}_{\mu\nu},\;\phi_a=\Omega^{-1}_{ab}\phi_b\;\Rightarrow\;g^{(a)}_{\mu\nu}=\frac{\phi^2_b}{\phi^2_a}g^{(b)}_{\mu\nu},\label{conf-t-ab}\end{align} link different physical states or gauges ``${\cal G}_a$'' and ``${\cal G}_b$,'' 

\begin{align} {\cal G}_a=(g^{(a)}_{\mu\nu},\phi_a),\;{\cal G}_b=(g^{(b)}_{\mu\nu},\phi_b)\;\Rightarrow\;{\cal G}_a\neq{\cal G}_b,\nonumber\end{align} which can not be identified. These are distinguished by different properties, such as: different proper time intervals $\Delta\tau_a=(\phi_b/\phi_a)\Delta\tau_b$, where $\Delta\tau_a=i\int ds_a$ ($ds_a^2=g^{(a)}_{\mu\nu}dx^\mu dx^\nu$) -- in general, different measuring procedure, different magnitude of the curvature invariants: $\hat R^{(a)}_{\lambda\mu\sigma\nu}\hat R_{(a)}^{\lambda\mu\sigma\nu}$, $\hat R^{(a)}_{\mu\nu}\hat R_{(a)}^{\mu\nu}$, $\hat R$, etc., different magnitude of the measured gravitational redshift, and so on. 

%-------------------2nd postulate-------------------

The second postulate on which the AAP rests: ``neither the masses of fields, nor the dimensionful constants, break WGS,'' is contrary to the CAP's assumption that dimensionful constants are not allowed, since these break WGS \cite{smolin-1979, cheng, bars, bars-sailing, ghilen-prd, ghilen-1, ghilen-2}. The former postulate is inspired by the approach to conformal transformations of the metric undertaken in \cite{dicke-1962}, within the context of Brans-Dicke (BD) theory of gravity. According to Dicke the conformal transformations of the metric in \eqref{gauge-t} are to be interpreted as transformations of the physical units. This approach is radically different from the assumption that masses and dimensionful constants, in general, do not transform under the WGT. According to \cite{dicke-1962} the masses should transform under \eqref{gauge-t} as: $m\rightarrow\Omega^{-1}m$. This transformation of mass is required by dimensional analysis and, besides, it is compatible with the property of $\tilde W_4$ space, that the length of vectors vary from point to point in space. In particular the mass squared, being (the negative of) the length squared of the four-momentum: $m^2=-g_{\mu\nu}p^\mu p^\nu$, should depend on spacetime point (see appendix \ref{app-c}.)

As long as we know, there is not any fundamental principle or law which fixes the way in which given dimensionful or dimensionless constants, should transform under the gauge transformations. Therefore, there is room for alternative assumptions on the transformation properties of given quantities. Notice that, if under \eqref{gauge-t} the masses transform as: $m\rightarrow\Omega^{-1}m$, then the mass term \eqref{sb-t}, does not break the WGS.\footnote{See the related discussion in section III.A.2 of Ref. \cite{faraoni-2007}.}

Since standard model (SM) fields acquire masses through the Higgs mechanism, one might wonder how can SM particles acquire point-dependent masses? The answer can be found in \cite{bars}, where it is shown that the SM can be modified in such a way as to be compatible with WGS. It is required to lift the mass parameter appearing in the quartic renormalizable potential, to a field with appropriate transformation properties under the gauge transformations: 

\begin{align} V(H,\phi)=\lambda_1\left(H^\dag H-\alpha^2\phi^2\right)^2+\lambda_2\phi^4,\label{higgs-pot}\end{align} where $\alpha$, $\lambda_1$ and $\lambda_2$ are dimensionless constants, $H$ is the Higgs field, which under \eqref{gauge-t} transforms as follows: $H\rightarrow\Omega^{-1}H,$ $H^\dag\rightarrow\Omega^{-1}H^\dag$. In \cite{bars} the field $\phi$ is identified with the dilaton and Riemann spaces are considered. If consider, instead, $\tilde W_4$ space, then $\phi$ must be identified with the Weyl gauge scalar of $\tilde W_4$ space. The vacuum expectation value (VEV) of the Higgs: $|H|^2:=H^\dag H=\alpha^2\phi^2$, is a point-dependent quantity. Means that the SM fields acquire point-dependent masses which under \eqref{gauge-t} transform like: $m\rightarrow\Omega^{-1}m$. 

As consequence of the AAP's second postulate it follows that Weyl scale invariance is not necessarily a broken symmetry. This entails, in turn, that acquirement of masses by the SM fields through $SU(2)\otimes U(1)$ symmetry breaking, does not imply breaking of WGS. Hence WGS may play an important role not only in short range and early time gravitational phenomena but, also in the present and future stages of the cosmic evolution.

%----------------WIG is not trivial---------------------

As stated in section \ref{sect-convent-app}, the condition $\der_{[\mu}Q_{\nu]}=0$, warrants that the SCE does not take place in $\tilde W_4$ space: I. e., that the length of given vector is not path-dependent. However, this condition does not forbid the length of vectors, being point-dependent \eqref{length-point-dep}. This and the simultaneous abandonment of the understanding of WGS as a mathematical redundancy, makes the apparently ``trivial'' $\tilde W_4$ spacetime to differ from $V_4$ space, where the length of vectors do not change during parallel transport.  

%---------------experiments--------------------

In what regards to experiments, local, point experiments are not able to determine whether WGS is a symmetry of the laws of physics, since the ``measuring stick'' undergoes the same spacetime variations as the quantity being measured. However, those experiments which imply comparison of quantities at different points in spacetime, such as the gravitational redshift, may be useful to test the WGS. This subject will be discussed in detail in section \ref{sect-z}.

%%%%%%%%%%%%%%%%%%%%%%%%%%%%%%%%%%%%%%%%%%%%%%%%%%%%%%%%%%%%%%%%%%%%%%%%%%%

\section{Weyl gauge invariant theory of gravity in $\tilde W_4$ space}
\label{sect-gtheor}

%%%%%%%%%%%%%%%%%%%%%%%%%%%%%%%%%%%%%%%%%%%%%%%%%%%%%%%%%%%%%%%%%%%%%%%%%%

Following the AAP proposed in this paper (see also \cite{quiros-2014, quiros-prd-2023-1, quiros-arxiv},) let us expose the basic elements of the gravitational theory depicted by the linear WGI gravitational Lagrangian \eqref{grav-lag}, which is based in $\tilde W_4$ space. In \cite{quiros-prd-2023-1} we have shown that the only WGI theory of gravity which is linear in the curvature scalar and admits coupling of matter fields with nonvanishing masses, is given by the gravitational Lagrangian: 

\begin{align} L_\text{grav}=\frac{\sqrt{-g}}{2}\left[\phi^2\hat R-\frac{\lambda}{4}\phi^4\right],\label{g-lag}\end{align} where $\lambda$ is a free constant. This theory is based in $\tilde W_4$ background spaces. Hence, in the above Lagrangian $\hat R=g^{\mu\nu}\hat R^\lambda_{\;\;\mu\lambda\nu}$ is the curvature scalar of $\tilde W_4$ space and $\phi$ is the Weyl gauge scalar determining the nonmetricity properties of WIG according to \eqref{grad-nm}. Thanks to the decomposition of WIG objects and operators in terms of their Riemannian equivalents, as for instance: $\hat R^\alpha_{\;\;\mu\beta\nu}=R^\alpha_{\;\;\mu\beta\nu}+\nabla_\beta L^\alpha_{\;\;\nu\mu}-\nabla_\nu L^\alpha_{\;\;\beta\mu}+L^\alpha_{\;\;\beta\lambda}L^\lambda_{\;\;\nu\mu}-L^\alpha_{\;\;\nu\lambda}L^\lambda_{\;\;\beta\mu}$, we have that: 

\begin{align} \hat R=R+6(\der\phi)^2/\phi^2-3(\nabla^2\phi^2)/\phi^2,\label{r-decomp}\end{align} where $\nabla^2\equiv g^{\mu\nu}\nabla_\mu\nabla_\nu$. The quantities and operators in the right-hand side (RHS) of \eqref{r-decomp}, are defined with respect to the LC connection $\{^\alpha_{\mu\nu}\}$. Hence, for instance, $R$ is the standard curvature scalar of Riemann space $V_4$ and $\nabla_\mu$ is the LC covariant derivative, etc.\footnote{Taking into account \eqref{r-decomp} -- up to a total derivative -- the gravitational Lagrangian \eqref{g-lag} can be written as in \eqref{grav-lag}. Hence, in terms of the composite (physical) metric of the CAP, the Lagrangian \eqref{g-lag} can be written as \eqref{triv-grav-lag}.}

%--------------------------EOM-----------------------------

Variation of the overall Lagrangian: 

\begin{align} L_\text{tot}=L_\text{grav}+L_\text{mat},\label{tot-lag}\end{align}  with respect to the metric, where $L_\text{grav}$ is given by \eqref{grav-lag} and $L_\text{mat}$ is the matter Lagrangian, leads to the following equation of motion (EOM):
    
\begin{align} &G_{\mu\nu}-\frac{1}{\phi^2}\left(\nabla_\mu\nabla_\nu-g_{\mu\nu}\nabla^2\right)\phi^2+\frac{\lambda}{8}\phi^2g_{\mu\nu}\nonumber\\
&+\frac{6}{\phi^2}\left[\der_\mu\phi\der_\nu\phi-\frac{1}{2}g_{\mu\nu}(\der\phi)^2\right]=\frac{1}{\phi^2}T^{(\text{mat})}_{\mu\nu},\label{grav-eom}\end{align} where $T^{(\text{mat})}_{\mu\nu}=-(2/\sqrt{-g})(\delta L_\text{mat}/\delta g^{\mu\nu}),$ is the stress-energy tensor of matter. Meanwhile, variation of $L_\text{tot}$ with respect to the scalar field yields the ``Klein-Gordon'' EOM:

\bea R+6\frac{(\der\phi)^2}{\phi^2}-3\frac{\nabla^2\phi^2}{\phi^2}-\frac{\lambda}{2}\phi^2=-\frac{1}{\phi^2}T^{(\text{mat})}.\label{phi-eom}\eea While deriving the latter equation we have taken into account that under WGS, variation with respect to the metric and with respect to the gauge field, are not independent of each other, since these obey \cite{quiros-prd-2023-1} (see Eq. (3) of Ref. \cite{jackiw-2014}):

\begin{align} \delta g_{\mu\nu}=-2\frac{\delta\phi}{\phi}g_{\mu\nu},\;\delta g^{\mu\nu}=2\frac{\delta\phi}{\phi}g^{\mu\nu}.\nonumber\end{align} These relationships, which lead to Eq. \eqref{phi-eom}, take place in the WGI theory of gravity over $\tilde W_4$ space, as well as, over $V_4$. However, in the latter case only matter with vanishing trace of the stress-energy tensor, $T^{(\text{mat})}=0,$ can couple to gravity.\footnote{Recall that Riemann geometry is not WGI in general. Although null-geodesics in $V_4$ are indeed WGI, Riemannian time-like geodesics are transformed by \eqref{gauge-t}.} Hence, in the latter case an equation like \eqref{phi-eom}, does not arise.

Equation \eqref{phi-eom} is not an independent equation since it coincides with the trace of \eqref{grav-eom}. In consequence $\phi$ does not obey an specific EOM: it is a free function which can be chosen at will.\footnote{In Ref. \cite{wig1} this property of the gauge scalar, within the cosmological context, was coined as ``marionette universe.''} Given that $\phi$ is not a dynamical degree of freedom (DOF) the kinetic term for the scalar field in \eqref{grav-lag} does not affect the measured Newton's constant, as it does in BD type theories.\footnote{In the BD theory, since the scalar field $\phi$ is a dynamical DOF, the measured gravitational constant is modified not only by $\phi$, but also by the kinetic term through the BD coupling constant $\omega_\text{BD}$ \cite{bd-theory, fujii-book, quiros-rev}: $$8\pi G_N=\frac{1}{\phi^2}\left(\frac{4+2\omega_\text{BD}}{3+2\omega_\text{BD}}\right).$$} The measured Newton's constant in the class of theories \eqref{grav-lag} corresponds to the tensor gravitational force. It is given by: $8\pi G_N(x)=M^{-2}_\text{pl}(x)=\phi^{-2}(x),$ where $M^2_\text{pl}(x)$ is the point-dependent squared effective Planck mass. For the same reason that $\phi$ is not a dynamical DOF, the wrong sign of the kinetic energy density term for $\phi$ in $L_\text{grav}$, is harmless.

Since the scalar field $\phi$ in equation \eqref{grav-lag} is not uniquely determined by the EOM, the metric is determined up to a conformal equivalence class: $g_{\mu\nu}\rightarrow\Omega^2 g_{\mu\nu}$, $\phi\rightarrow\Omega^{-1}\phi$. Each choice $\phi=\phi(x)$, picks out a gauge. Since, in line with the AAP's first postulate, WGS is not assumed as a redundancy, the different gauges are not identified so that these represent different physical states which can be experimentally differentiated.\footnote{According to the AAP, WGS is not properly a symmetry in the conventional sense. This result is consistent with the finding that the Noether current associated with WGS, vanishes \cite{jackiw-2014, oda-2022, rodrigo-arxiv}.}

%========================================================

\subsection{Physical significance of gauge freedom}
\label{sect-gauges}

%========================================================

The geometric scalar field $\phi$, being a free function, can be any smooth function $\phi=\phi_a(x)$, where $a=0,1,2,...,N$ ($N\rightarrow\infty$,) labels the different functions. Means that each choice of a function $\phi_a$ or, properly a gauge choice, leads to a specific theory with its own set of measured quantities. Due to the AAP's first postulate stating that WGS is not a mathematical redundancy, we can not identify the different gauges.

The general element of a gauge can be defined in the following way \cite{quiros-prd-2023-1}:

\bea {\cal G}_a:=\{({\cal M}_4,g^{(a)}_{\mu\nu},\phi_a)|L^{(a)}_\text{tot},{\cal S}_a,{\cal C},\cdots\},\label{gauge-a}\eea where ${\cal M}_4\in\tilde W_4$ is a WIG spacetime manifold. In this expression $L^{(a)}_\text{tot}=L^{(a)}_\text{grav}+L^{(a)}_\text{sm}$, where $L^{(a)}_\text{grav}$ is the gravitational Lagrangian \eqref{grav-lag} in the gauge ${\cal G}_a$:

\bea L^{(a)}_\text{grav}=\frac{\sqrt{-g}}{2}\left[\phi_a^2 R_a+6(\der\phi_a)^2-\frac{\lambda}{4}\phi_a^4\right],\label{gauge-a-lag}\eea with $R_a\equiv R[g^{(a)}_{\mu\nu}]$, and $L^{(a)}_\text{sm}$ is the SM Lagrangian, including the modified gauge invariant Higgs Lagrangian \cite{bars}, specialized to ${\cal G}_a$. In equation \eqref{gauge-a}, ${\cal S}_a$ represents the set of measured quantities (for instance, the redshift,) ${\cal C}$ is the set of gauge invariant constants (the Planck constant $\hbar$, the speed of light in vacuum $c$ and the quantum of electric charge $e$, among others,) while the ellipsis represent other relevant measured quantities. 

Due to WGS the collection of all possible gauges forms an equivalence class of conformally related theories, instead of a single theory. This conformal equivalence class can be expressed as \cite{quiros-prd-2023-1}:

\bea {\cal K}=\left\{{\cal G}_0,{\cal G}_1,{\cal G}_2,...,{\cal G}_a,...,{\cal G}_N|\;a\in\mathbb{N}\right\},\label{ce-class}\eea where $N\rightarrow\infty$. The AAP's first postulate warrants that ${\cal G}_i\neq{\cal G}_j$ for $i\neq j$, where the subindices $i$ and $j$ independently run from $0$ to $N$ ($i,j=0,1,...,N$.) Any two different gauges ${\cal G}_a$ and ${\cal G}_b$, in ${\cal K}$, are related by the conformal transformation \eqref{conf-t-ab}: $g^{(a)}_{\mu\nu}=\left(\phi_b/\phi_a\right)^2g^{(b)}_{\mu\nu}$. Under the latter transformation the gravitational Lagrangian $L^{(a)}_\text{grav}$, transforms into $L^{(b)}_\text{grav}=\sqrt{-g}[\phi_b^2 R_b+6(\der\phi_b)^2-\lambda\phi_b^4/4]/2,$ while Weyl integrable geometry space transforms into itself: $\tilde W_4\rightarrow\tilde W_4$. Means that the gravitational laws are invariant under \eqref{gauge-t}: The laws of gravity look the same in any gauge, but for the GR gauge where these take the (simplest) GR form.\footnote{The fact that the GR laws are not Weyl gauge invariant only means that a choice of gauge breaks the manifest WGS of the theory. The same is true, for instance, when one chooses a specific set of coordinates: the explicit general coordinate invariance of the GR theory is lost.} According to the AAP, the transformation of units $g^{(a)}_{\mu\nu}=\Omega^2_{ab}g^{(b)}_{\mu\nu}$ in Eq. \eqref{conf-t-ab}, must be interpreted as a linkage between two different physical states.

In the present case the laws of gravity are the same in any gauge, however, the geometrical and physical properties of background space change from gauge to gauge. This is to be contrasted with the conventional approach to WGS, according to which the same theory, depicted by equations \eqref{tot-lag}, \eqref{grav-eom} over $\tilde W_4$ space, is fully equivalent to GR theory over $V_4$ space and where the Weyl gauge transformations \eqref{gauge-t} collapse to the identity transformation (no transformation at all.)

%--------------------------GR gauge--------------------------

\subsubsection{General relativity gauge}\label{subsect-grg}

%------------------------------------------------------------

An outstanding gauge in ${\cal K}$, which we identify as ${\cal G}_0$, is the one obtained if in \eqref{grav-lag} we set $\phi(x)=\phi_0=$ const. Since it coincides with general relativity, we shall call this as ``GR gauge.'' ${\cal G}_0$ can be obtained as well from any gauge ${\cal G}_a\in{\cal K}$ through the gauge transformation \eqref{conf-t-ab}: $g^{(a)}_{\mu\nu}=\Omega^2_{a0}g^{(0)}_{\mu\nu}$, $\phi_a=\Omega^{-1}_{a0}\phi_0,$ or

\begin{align} \;g^{(a)}_{\mu\nu}=\left(\frac{M_\text{pl}}{\phi_a}\right)^2g^\text{gr}_{\mu\nu},\label{gauge-t-a0}\end{align} where we made the following identifications: $g^\text{gr}_{\mu\nu}\equiv g^{(0)}_{\mu\nu}$ and $\phi_0\equiv M_\text{pl}$ -- the constant Plack mass. Under \eqref{gauge-t-a0}, the Lagrangian \eqref{gauge-a-lag} is transformed into the Einstein-Hilbert (EH) Lagrangian:

\bea L_\text{eh}=\frac{\sqrt{-g}M^2_\text{pl}}{2}\left(R_\text{gr}-2\Lambda\right),\label{eh-lag}\eea where $R_\text{gr}\equiv R[g^\text{gr}_{\mu\nu}]$ and $\Lambda=\lambda M^2_\text{pl}/8$ is the cosmological constant. The inverse transformation, if it exists, maps the EH Lagrangian $L_\text{eh}$, back into the Lagrangian $L^{(a)}_\text{grav}$. Under \eqref{gauge-t-a0} $\tilde W_4$ is transformed into Riemann space: $\tilde W_4\rightarrow V_4$. Hence, the GR gauge ${\cal G}_0$ amounts to standard GR over $V_4$ background space. Means that GR is just a gauge in the conformal equivalence class defined in \eqref{ce-class}. Although GR itself is not manifest gauge invariant, it belongs in a larger class ${\cal K}$ of WGI gravitational theories.

%================================================

\subsection{On the role of the gauge invariants}
\label{sect-ginv}

%================================================

Due to the AAP's first postulate the different gauges may not be identified, so that the gauge invariant quantities are not the only physically meaningful quantities. The invariants of the geometry, such as the curvature scalar, the Kretschmann invariant, etc., also have the physical meaning, just as in theories which are not WGI. In consequence, one may wonder which is the role of the gauge invariants according to the AAP? 

The following diffeomorphism invariants of Weyl integrable geometry: the Kretschmann invariant $\hat K:=\hat R^{\sigma\mu\lambda\nu}\hat R_{\sigma\mu\lambda\nu}$, the product $\hat L:=\hat R_{\mu\nu}\hat R^{\mu\nu}$, the curvature scalar $\hat R:=g^{\mu\nu}\hat R_{\mu\nu}$, etc., are not WGI. In contrast, the quantities: $\phi^{-4}\hat K$, $\phi^{-4}\hat L$ and $\phi^{-2}\hat R$, are not only invariant under general coordinate transformations, but these are also invariant under the WGT \eqref{gauge-t}. Hence, if we know, for instance, the GR static, spherically symmetric Kretschmann scalar: $K_\text{gr}=48m^2/r^6$, we have that (we chose the normalization $\phi_\text{gr}=M_\text{pl}=1$): $\phi^{-4}(r)\hat K(r)=K_\text{gr}=48m^2/r^6$, which leads to $\hat K(r)=48m^2\phi^4(r)/r^6.$ Hence, different choices of the free function $\phi(r)$, define different gauges with given expressions for the Kretschmann scalar. The same is true for the other gauge invariant quantities. The gauge invariants allow us to relate physical quantities in a given gauge with the same quantities in a different gauge.

%------------------------FRW example---------------------

Another illustration can be based in a cosmological example. Consider the Friedmann-Robertson-Walker (FRW) line element with flat spatial sections: $ds^2=-dt^2+a^2(t)\delta_{ik}dx^idx^k$, where $t$ is the cosmic time and $a(t)$ is the dimensionless, time-dependent scale factor. In this case, the EOM \eqref{grav-eom}, plus the continuity equation: $\nabla^\mu T^{(r)}_{\mu\nu}=0,$ amount to the following independent differential equations \cite{quiros-prd-2023-1}:

\begin{align} 3\left(\frac{a'}{a^2}+\frac{\phi'}{a\phi}\right)^2=\frac{1}{\phi^2}\rho_r,\;\rho'_r+4\frac{a'}{a}\rho_r=0,\label{frw-eom}\end{align} where the tilde means derivative with respect to the conformal time $\tau=\int dt/a(t)$ and, for simplicity, we have considered radiation with energy density $\rho_r$ as the background matter.\footnote{The advantage of considering the conformal time $\tau$ instead of $t$ relies in the fact that the former -- being a coordinate time -- is not modified by the gauge transformations \eqref{gauge-t}.} Straightforward integration of the right-hand equation in \eqref{frw-eom} yields: $\rho_r=\mu_0^2/a^4$, where $\mu_0$ is an integration constant. Then, if introduce the gauge invariant variable $v\equiv\phi a$, the left-hand equation in \eqref{frw-eom} can be written in the following manifest WGI way: $v'=\mu^2_0/\sqrt{3}$. Integration of this equation leads to the gauge invariant expression: $v(\tau)=\phi(\tau)a(\tau)=\mu^2_0\tau/\sqrt{3},$ where, without loss of generality, we have considered a vanishing integration constant. There are not other independent differential equations in the gravitational EOM, so that most we can determine, after appropriate mathematical handling of the equations of motion, is the gauge invariant combination $v=\phi a$. Different choices, either of the free function $\phi(\tau)$ or of the scale factor $a(\tau)$, define different gauges. Let us assume, for instance, that $a_\text{gr}(\tau)$ is a known solution of the cosmological GR EOM. Then, since $v$ is a gauge invariant, we have that: $v(\tau)=\phi(\tau)a(\tau)=a_\text{gr}(\tau)$, where we assumed that $\phi_\text{gr}=1$. From this equation it follows that $a(\tau)=\phi^{-1}(\tau)a_\text{gr}(\tau)$. Hence, different choices of $\phi(\tau)$ generate different behaviors of the scale factor which are conformal to that of GR.

%%%%%%%%%%%%%%%%%%%%%%%%%%%%%%%%%%%%%%%%%%%%%%%%%

\section{Redshift effect in $\tilde W_4$ space}
\label{sect-z}

%%%%%%%%%%%%%%%%%%%%%%%%%%%%%%%%%%%%%%%%%%%%%%%%%

According to the CAP, in terms of the physical metric, the WGT collapse to the identity transformation and the gravitational theory \eqref{grav-lag} over $\tilde W_4$ space transforms into plain GR. Therefore, the WGI gravitational theory \eqref{grav-lag} over $\tilde W_4$ space, does not have observational consequences beyond those of GR theory. This is to be contrasted with the AAP, according to which the WGI theory \eqref{grav-lag} over $\tilde W_4$ space bears measurable physical consequences, so that it can be observationally tested. For this reason here we follow the AAP.

%--------------------------------------------------

A relevant feature of $\tilde W_4$ space, is that the length of vectors varies from point to point in spacetime \cite{weyl-1918, weyl-book, london-1927, dirac-1973, utiyama-1973, maeder-1978, many-weyl-book, wig2, wig4, wig10, wig11}. Since the units of measure change in the same way as the measured quantity does, by means of local, point experiments, there is no way in which the measurements performed in one gauge can be differentiated from measurements in a different gauge. Nevertheless, measurements which imply comparison of quantities evaluated at different spacetime points may differentiate between gauges. One such measurement is the gravitational shift of frequency. In this case a photon with certain frequency is emitted at some point with a given value of the gravitational potential and it is then detected at some other point with a different value of the gravitational potential. The measured frequency of the photon at emission is then compared with its measured frequency at detection.

There are two different sources of frequency shift in $\tilde W_4$ space: 1) the standard curvature shift which is due to the influence of the curvature of background space on the propagation of photons, and 2) the shift of frequency due to the variation of atomic transition energies in spacetime, which we call as ``nonmetricity'' shift of frequency. While in the GR gauge ${\cal G}_0$ over $V_4$, only the first kind of frequency shift arises, in any other gauge ${\cal G}_a$, both types of shift occur.

%=================================

\subsection{Curvature redshift}

%=================================

Let us start by describing the gravitational redshift in the GR gauge. For definiteness we assume that in ${\cal G}_0$ the photon propagates in the radial direction in a static, spherically symmetric Schwarzschild spacetime with metric: 

\bea ds_{(0)}^2=-A^2dt^2+A^{-2}dr^2+r^2d\Omega^2,\label{0-line-e}\eea where $A^2=-g^{(0)}_{00}=1-2m/r$. Let us further assume that the photon follows the radial direction with angular coordinates $(\theta,\vphi)=(\pi/2,0)$. Then, the components of the wave-vector of the photon in ${\cal G}_0$, read: 

\bea k_{(0)}^\mu\equiv\frac{dx^\mu}{d\xi_{(0)}}\;\Rightarrow\;k_{(0)}^\mu=\left(\omega_{(0)}, k_{r(0)},0,0\right),\label{wave-v-comp}\eea where $\omega_{(0)}=dt/d\xi_{(0)}$ and $k_{r(0)}=dr/d\xi_{(0)}$. Besides $k^{(0)}_\mu k_{(0)}^\mu=-A^2\omega_{(0)}^2+A^{-2}k^2_{r(0)}=0$. The null geodesics \eqref{0-geod} reduce to the following equations: 

\bea \frac{d\omega_{(0)}}{\omega_{(0)}}=-2\frac{dA}{A},\;\frac{dk_{r(0)}}{d\xi_{(0)}}=0,\label{0-geod'}\eea where we have taken into account that $k_{r(0)}dA/dr=dA/d\xi_{(0)}$. Straightforward integration of the first equation above leads to: $\omega_{(0)}=\omega_0/A^2=\omega_0/(-g^{(0)}_{00})$, where $\omega_0$ is an integration constant. The physical cyclic frequency reads: $\omega^{(0)}_\text{ph}=\sqrt{-g^{(0)}_{00}}\omega_{(0)}$, hence:

\bea \omega^{(0)}_\text{ph}(r)=\frac{\omega_0}{\sqrt{-g^{(0)}_{00}(r)}}=\frac{\omega_0}{\sqrt{1-2m/r}},\label{phys-w-0}\eea which is the standard textbook result.

Suppose that a photon is emitted at some time $t$ by an hydrogen atom placed at spatial point $(r,\pi/2,0)$, and it is then absorbed at a later time $t_0$ by an identical hydrogen atom placed at $(r_0,\pi/2,0)$. Due to the effect of curvature on the propagation of the photon, there will be a relative shift of photon's frequency: 

\bea z^{(0)}_\text{curv}=\frac{\nu_{(0)}^\text{em}-\nu_{(0)}^\text{abs}}{\nu_{(0)}^\text{abs}}=\frac{\nu_{(0)}^\text{em}}{\nu_{(0)}^\text{abs}}-1,\label{z-curv-gen}\eea where $\nu_{(0)}^\text{em}\equiv\omega^{(0)}_\text{ph}(r)/2\pi$ is the measured frequency of the emitted photon, $\nu_{(0)}^\text{abs}\equiv\omega^{(0)}_\text{ph}(r_0)/2\pi$ is its frequency when it is absorbed by the second atom and $z^{(0)}_\text{curv}$ is the relative (curvature) ``redshift'' of frequency in the GR gauge.\footnote{The photon is absorbed by the second hydrogen atom only if it is placed in a centrifuge with controlled rotation speed, as in Mossbauer experiment, so that the gravitational shift of frequency is compensated by the Doppler shift.} In the GR gauge this is the only source of shift of photon's frequency. 

Taking into account equations \eqref{phys-w-0} and \eqref{z-curv-gen}, the redshift of frequency in ${\cal G}_0$ is given by (compare with Eq. (6.3.5) of Ref. \cite{wald-book} or with Eq. (9.12) of Ref. \cite{bambi-book}):

\bea z^{(0)}_\text{curv}=\frac{\sqrt{-g^{(0)}_{00}(r_0)}}{\sqrt{-g^{(0)}_{00}(r)}}-1.\label{z-curv-0}\eea Since the Riemannian null-geodesic equations \eqref{0-geod} are not modified by the gauge transformations \eqref{gauge-t}, then the same curvature shift of frequency arises in any other gauge ${\cal G}_a$:\footnote{Notice that, acording to the CAP, since the physical metric is $\mathfrak{g}_{\mu\nu}=\phi^2g_{\mu\nu}$, then $$z^{(a)}_\text{curv}=\sqrt\frac{-\phi_a^2(r_0)g^{(a)}_{00}(r_0)}{-\phi_a^2(r)g^{(a)}_{00}(r)}-1=\sqrt\frac{-g^{(0)}_{00}(r_0)}{-g^{(0)}_{00}(r)}-1=z^{(0)}_\text{curv},$$ so that the gravitational redshift is the same in any gauge.}

\bea z^{(a)}_\text{curv}=\frac{\sqrt{-g^{(a)}_{00}(r_0)}}{\sqrt{-g^{(a)}_{00}(r)}}-1.\label{z-curv-a}\eea 

Combining equations \eqref{gauge-t-a0}, \eqref{z-curv-0} and \eqref{z-curv-a}, the following relationship between the curvature redshift in the GR gauge ${\cal G}_0$ and the same measured quantity in any other gauge ${\cal G}_a$, is obtained:

\bea z_\text{curv}=\frac{\phi(r)}{\phi(r_0)}z_\text{gr}+\frac{\phi(r)}{\phi(r_0)}-1,\label{z-rel}\eea where we dropped the index ``$a$,'' which denotes a specific gauge and, also, we renamed the curvature redshift taking place in the GR theory: $z_\text{gr}\equiv z^{(0)}_\text{curv}$.

%==================================

\subsection{Nonmetricity redshift}

%==================================

Time-like particles follow autoparallels which coincide with the geodesics of $\tilde W_4$ space (see in appendices \ref{app-b} and \ref{app-c}.) The autoparallel curves satisfy \eqref{time-auto-p}. Meanwhile, in $\tilde W_4$ space, since the mass is a point-dependent quantity, then $m=m(x)$ can not be taken out of the action integral: $S=\int mds.$ From this action the time-like geodesic equation 

\begin{align} \frac{d^2x^\alpha}{ds^2}+\left\{^\alpha_{\mu\nu}\right\}\frac{dx^\mu}{ds}\frac{dx^\nu}{ds}-\frac{\der_\mu m}{m}h^{\mu\alpha}=0,\label{w4-time-geod}\end{align} is derived. Hence, since time-like autoparallels \eqref{time-auto-p} and time-like geodesics \eqref{w4-time-geod} coincide in $\tilde W_4$ space, we get that:\footnote{An independent mathematical derivation of \eqref{mass-autop} is given in appendix \ref{app-c}.} 

\bea \frac{dm}{m}=\der_\mu\ln\phi dx^\mu\;\Rightarrow\;\frac{\der_\mu m}{m}=\frac{\der_\mu\phi}{\phi}.\label{mass-autop}\eea This equation can be readily integrated to get:

\bea m(r)=m_0\frac{\phi(r)}{\phi(r_0)},\label{mass-phi-rel}\eea where $m_0=m(r_0)$ is the magnitude of the mass evaluated at the value $r_0$ of the radial coordinate. Equation \eqref{mass-phi-rel} expresses how much the mass of given time-like particle varies form point to point in $\tilde W_4$ spacetime. This would lead, in particular, to a change in the atomic transition energies in spacetime. 

According to \eqref{mass-phi-rel} the masses of particles, for instance of the electron: $m_e$, vary from point to point in spacetime: $m_e(r)=m_{e0}\phi(r)/\phi(r_0)$, where $m_{e0}$ is the value of the electron's mass at some reference point with spatial coordinates: $\vec{x}_0=(r_0,\pi/2,0)$. Let $\nu_{if}$ be the frequency of a photon emitted by an hydrogen atom located at some spatial point with coordinates $\vec{x}=(r,\pi/2,0)$, due to a transition from a state with principal quantum number $n_i$ into a state with $n_f$:

\bea \nu_{if}(r)=\frac{m_e(r)\alpha^2}{2}\left|\frac{1}{n_f^2}-\frac{1}{n_i^2}\right|,\label{nu-x}\eea where $\alpha$ is the fine structure constant. The frequency of the similar photon emitted/absorbed by an hydrogen atom placed at the reference point $\vec{x}_0$ is given by: $\nu_{if}(r_0)=m_{e0}\alpha^2|n_f^{-2}-n_i^{-2}|/2$. We further assume that a photon emitted by an hydrogen atom placed at $\vec{x}$, with frequency $\nu^\text{em}_{if}=\nu_{if}(r)$, is then absorbed by an hydrogen atom placed at $\vec{x}_0$. In addition to the modification of the photon's frequency during its propagation in spacetime, there is a shift of frequency $z_\text{nm}$ which is associated, exclusively, with variation of the atomic transition energies in spacetime:

\bea z_\text{nm}=\frac{\nu^\text{em}_{if}}{\nu^\text{abs}_{if}}-1=\frac{m_e(r)}{m_{e0}}-1=\frac{\phi(r)}{\phi(r_0)}-1,\label{z-nm}\eea where $\nu^\text{abs}_{if}=\nu_{if}(r_0)$. This shift of frequency is due to the nonmetricity of $\tilde W_4$ space. In Riemann space $V_4$, since $\phi(r)=\phi_0$ is a constant, then: $z_\text{nm}=0$. 

In $\tilde W_4$ spacetime the overall redshift is a sum of the curvature and of the nonmetricity redshifts: $z_\text{tot}=z_\text{curv}+z_\text{nm}$, or if take into account \eqref{z-rel}, we can write the overall redshift of frequency in terms of the curvature redshift $z_\text{gr}$ taking place in the GR gauge:

\bea z_\text{tot}=\frac{\phi(r)}{\phi(r_0)}z_\text{gr}+2\left[\frac{\phi(r)}{\phi(r_0)}-1\right].\label{z-tot}\eea This equation is independent of the type of atom which emits/absorbs the photon.

%====================================================

\subsection{Experimental differentiation of gauges}

%====================================================

Equation \eqref{z-tot} relates the measured gravitational redshift in the GR gauge with the measured redshift in any other gauge in ${\cal K}$. This means that we can experimentally differentiate the gauges through redshift experiments \cite{pound-rebka, pound-snider, vessot, will-lrr}.

For definiteness and in order to make the discussion more transparent, let us assume that the measurement of the redshift is performed in a Pound-Rebka type experiment \cite{pound-rebka, pound-snider}: A target $T_h$ made of some atom is placed at some height $h$ on top of a tower over the Earth surface (the radial coordinate $r=R_\oplus+h$, where $R_\oplus$ is the mean radius of the Earth) and an identical target $T_0$ (made of same atom) is placed at the bottom of the tower (radial coordinate $r_0=R_\oplus$.) Atoms in $T_h$ are in some excited state, so hat these are able to emit photons to return to the ground state. Emitted photons are then absorbed by atoms in the target $T_0$ which is located at the bottom of the tower.\footnote{Here we omit experimental details such as, for instance, that the target at the bottom of the tower must be mounted in a centrifuge with controlled rotation speed, etc.}

%-----------------------------------------

\subsubsection{General relativity gauge}

%-----------------------------------------

The gravitational redshift measured in the GR gauge ${\cal G}_0$ is given by \eqref{z-curv-0}:

\bea z_\text{gr}=\frac{\sqrt{1-\frac{2m}{R_\oplus}}}{\sqrt{1-\frac{2m}{R_\oplus+h}}}-1\approx-\frac{gh}{c^2},\label{z-gr}\eea where we took into account that $m\equiv G_NM_\oplus/c^2$ ($M_\oplus$ is the mass of the Earth), $g\equiv G_NM_\oplus/R^2_\oplus$ and we assumed the linear approximation. Hence, if choose the following experimental values: $h=1.00\times 10^2$m, $R_\oplus=6.371\times 10^6$m, $g=9.807$m/s$^2$ and $c=2.998\times 10^8$m/s, the magnitude of the gravitational redshift in general relativity is: $|z_\text{gr}|\approx 1.091\times 10^{-14}$.

%--------------------------------

\subsubsection{Arbitrary gauge}

%--------------------------------

Let us compute the quantity $\phi(r)/\phi(r_0)$ in another (arbitrary) gauge. Since $\phi(r)$ is a free function, we can fix it to be any continuous function. Let us assume, for definiteness, the following bi-parametric class of functions:

\begin{align} \phi(r)=\sqrt{1+k\left(\frac{2m}{r}\right)^{2n}},\label{phi-choice}\end{align} where $k$ is a dimensionless real constant and $n$ is a positive real. In consequence we have that:

\begin{align} \frac{\phi(r)}{\phi(r_0)}&=\frac{\sqrt{1+k\left(\frac{2m}{R_\oplus+h}\right)^{2n}}}{\sqrt{1+k\left(\frac{2m}{R_\oplus}\right)^{2n}}}\nonumber\\
&\approx 1-4^nknz_\text{gr}^{2n}\left(\frac{R_\oplus}{h}\right)^{2n-1},\label{z-phi-g2}\end{align} where $z_\text{gr}$ is given by \eqref{z-gr}, so that $|z_\text{gr}|\approx 1.091\times 10^{-14}$. If we substitute \eqref{z-phi-g2} back into \eqref{z-tot} in order to get the magnitude of the redshift measured in the arbitrary gauge, we obtain: $z_\text{tot}=(1+\alpha)z_\text{gr},$ where 

\begin{align} |\alpha|\approx 2^{2n+1}kn\left(\frac{R_\oplus}{h}|z_\text{gr}|\right)^{2n-1}.\label{alpha}\end{align} For the chosen value of the height of the experimental tower ($h=100$m,) we get that $|z_\text{gr}|R_\oplus/h\approx 0.7\times 10^{-9}$. Meanwhile, the most precise standard redshift test to date yields the limit $|\alpha|<2\times 10^{-4}$ \cite{vessot, will-lrr}. Hence, for $n>13/18$, this limit is satisfied, where in order to get a definite estimate, we assumed that $2^{2n+1}nk\sim 1.$ In this case, by means of redshift experiments, it is not possible to differentiate the GR gauge from an arbitrary gauge with $\phi(r)$ given by \eqref{phi-choice}.

%%%%%%%%%%%%%%%%%%%%%%%%%%%%%%%%%%%%%%%%%%%%%%%%%%%%%%%%%%%%%%%%%%

\section{Critique on the Alternative Approach to Gauge Symmetry}
\label{sect-critique}

%%%%%%%%%%%%%%%%%%%%%%%%%%%%%%%%%%%%%%%%%%%%%%%%%%%%%%%%%%%%%%%%%%

During the years, while we were developing the AAP exposed in this and in other papers \cite{quiros-2014, quiros-prd-2023-1, quiros-arxiv}, we have received continous critique on our work, that arises from ignorance of the postulates under which a given formalism or approach is based. Most times it is implicitly assumed that the CAP is correct and, besides, that it is the only possible understanding of WGS. In some of the comments, the discussion about the conformal transformations and the different frames in which scalar-tensor theories of gravity can be formulated, is confounded with the Weyl scale transformations and the WGS issue. We think that discussion of these critical remarks will be of help to correctly understand the reach of our approach. Below we list some of the critical comments on our unconventional understanding of WGS and, at the end, we address each one of these items.

\begin{enumerate}

%---------------------------------------

\item In theories with scale symmetry dimensionful couplings are forbidden.

\item The appearance of a mass parameter is confusing. Such a dimensionful quantity cannot exist in a Weyl-invariant theory and instead only couplings to other fields, which also transform under gauge transformations, are allowed.

\item The AAP and its physical consequences do not represent relevant new contribution to this subject. It is known that GR can be viewed as a gauge fixed version of a Weyl invariant theory of a conformal class of metrics and a conformally coupled scalar. This is obviously true because physics does not depend on local choices of unit systems.

\item Lagrangian \eqref{grav-lag} is trivial since the Weyl vector, being only a gradient of a scalar field, can be gauged away.

\item A conformal transformation of the action \eqref{grav-lag} into the Einstein frame would show that the scalar field disappears from the action completely. The theory is thus equivalent to the usual metric formulation of gravity with the Einstein-Hilbert action. There are no extra physical effects.

%\item Regarding the choice of gauge and physically meaningful quantities let us consider, for instance, the fermionic density $\bar{\Psi}\Psi$. The only scale-invariant contribution to the action one can form from it is $\phi\bar{\Psi}\Psi$. A gauge choice cannot have observable consequences, so that, for instance, the redshift measurements are gauge-independent.

\item The AAP leads to conclude that the different frames correspond to different physics, however, the change of frame is a coordinate transformation in the space of fields. If invertible, it cannot change the classical physics. See for example the frame-covariant formalism in \cite{comment-3}.

\item The question of what is the "physical frame" relates to the coupling to matter, i.e. in which frame (if any) the observable fields are minimally coupled to the metric. If the matter coupling is also properly transformed, even this semantic issue goes away. Then the physics is frame independent, as noted in \cite{thooft-2015} (see also \cite{comment-4}.)

\item The "fifth force" that would arise upon a conformal transformation of timelike geodesics, is an apparent effect. A true matter particle becomes coupled to the scalar field after the conformal transformation so it does not fall along the transformed geodesics.

%---------------------------------------------------------------
    
\end{enumerate}

%---------------------items 1 and 2-----------------

The comment in the first and second items is one of the most frequent critique on our approach to WGS. It is based in the understanding that the CAP is the only possible approach to Weyl gauge symmetry. Specifically, this comment follows from the CAP's second postulate. Our approach incorporates just the opposite postulate: ``neither the masses of fields, nor the dimensionful constants, break WGS,'' so that this critique is on the AAP's second postulate. However, a postulate is ``a thing suggested or assumed as true as the basis for reasoning.'' 

The AAP's second postulate originates from the absence of a fundamental principle or law of Nature that states the specific way in which dimensionful constants and quantities, in general, should transform under the WGT \eqref{gauge-t}. The only well-established specification on how given quantities should transform is, precisely, in the mathematical formulation of WGT \eqref{gauge-t}: the metric tensor should transform like $g_{\mu\nu}\rightarrow\Omega^2g_{\mu\nu}$, so that the line element and derived quantities should transform accordingly: $ds^2\rightarrow\Omega^2ds^2$, $\Gamma^\alpha_{\;\;\mu\nu}\rightarrow\Gamma^\alpha_{\;\;\mu\nu},$ etc. In particular, the proper time element $d\tau=\sqrt{-g_{00}}dt$, which is measured in time units $[T]$, and the proper length element $dl=\sqrt{g_{ik}dx^idx^k}$, which is measured in length units $[L]$, transform in the same way. Means that the units of proper time $[T]$ and of proper length $[L]$, transform under \eqref{gauge-t} as: $[T]\rightarrow\Omega[T]$ and $[L]\rightarrow\Omega[L]$, respectively. Similarly, the four-velocity vector $u^\mu=dx^\mu/d\tau$, is constrained to transform as $u^\mu\rightarrow\Omega^{-1}u^\mu$. That the EM vector potential $A_\mu$ is not transformed by \eqref{gauge-t}, is an additional postulate of the transformations of units in \cite{dicke-1962}. The latter property does not follow from a mathematical consistency requirement, but it is a physical consistency postulate. Other quantities, including the mass $m$, the Planck's constant $\hbar$ and the electron's charge $e$, are not obliged to transform in a specific way because of any principle or law. On the contrary, at hoc assumptions are required. For instance, $\hbar$ is measured in $[ML^2/T]$ units ([M] is the mass unit.) Hence, if one assumes that, under the WGT \eqref{gauge-t}, the mass is not transformed, the direct consequence is that the Planck's constant should transform as: $\hbar\rightarrow\Omega\hbar.$ Otherwise, if $\hbar$ does not transform under \eqref{gauge-t}, then the masses should transform as in \cite{dicke-1962}: $m\rightarrow\Omega^{-1}m.$ In this case the four-momentum $p^\mu=mu^\mu$, necessarily transforms in the following way: $p^\mu\rightarrow\Omega^{-2}p^\mu.$ If assume, additionally, that the electron's charge is unchanged by the WGT: $e\rightarrow e$, then physical consistency requires that the EM vector potential transforms like the four-momentum does: $A^\mu\rightarrow\Omega^{-2}A^\mu.$ As a consequence $A_\mu\rightarrow A_\mu.$ 

The learned lesson is that ad hoc assumptions are necessarily required. This is our rationale for the replacement of the second postulate on which the conventional approach to Weyl gauge symmetry is based. Besides, the AAp's second postulate finds physical basement in the modification of the Higgs Lagrangian proposed in \cite{bars}, where the mass parameter in the quartic renormalizable potential \eqref{higgs-pot}, is lifted to a field with appropriate transformation properties under the WGT \eqref{gauge-t}, so that it be compatible with WGS. The masses acquired by the SM fields after $SU(2)\otimes U(1)$ symmetry breaking are thus point dependent quantities, which under the WGT \eqref{gauge-t} transform like $m\rightarrow\Omega^{-1}m$, as required.

In the last part of the second item it is said that ``only couplings to other fields, which also transform under gauge transformations, are allowed.'' This is precisely what the Yukawa couplings of fermions to the Higgs field means. Recall that in the Higgs potential \eqref{higgs-pot} the mass parameter $v$ has been lifted to the category of a field \cite{bars}: $v\rightarrow\alpha\phi,$ where $\alpha$ is a dimensionless coupling and $\phi$ is the Weyl gauge field of $\tilde W_4$ space, i. e., it is the scalar field in the gravitational Lagrangian \eqref{grav-lag}. The mass $m_f$ acquired by given fermion field after $SU(2)\otimes U(1)$ symmetry breaking, is given by:

\begin{align} m_f=\frac{\alpha}{\sqrt{2}}g_f\phi,\nonumber\end{align} where $g_f$ is the Yukawa coupling of the Higgs to the fermion field. Hence, under \eqref{gauge-t}; $m_f\rightarrow\Omega^{-1}m_f,$ as required.

%------------------item 3-------------------------------

The third comment above is a very specific opinion. We included it because we want to recognize that our understanding of gauge freedom is very much like to the one undertaken in \cite{comment-2, shaukat-1, shaukat-2, shaukat-3}. However, in our works \cite{quiros-2014, quiros-prd-2023-1, quiros-arxiv}, including the present paper, we follow a completely different way. We show, in particular, that the gauge choice has physical consequences, which can be observationally tested. As long as we know, but for the above references, there are no other works in the bibliography where a similar approach to the Weyl gauge transformations, is undertaken. Besides, the last statement in the third item is the same as Dicke's postulate \cite{dicke-1962}. To our knowledge, because of the CAP's second postulate, there is no place for WGS in the phenomenology after $SU(2)\otimes U(1)$ symmetry breaking and the consequent acquirement of mass by the SM fields. Hence, no matter how obvious it may seem that ``physics does not depend on local choices of unit systems,'' this postulate has not been incorporated into physical theories yet (with the exception of references \cite{comment-2, shaukat-1, shaukat-2, shaukat-3} and our own work \cite{quiros-2014, quiros-prd-2023-1, quiros-arxiv}.) 

%-------------------items 4 and 5-----------------------

The fourth and fifth comments are consequence of the implicit assumption that the CAP is the only possible approach to WGS. In particular these comments are based on the CAP's first postulate. Our approach is based on quite the opposite postulate. Our rationale for replacing CAP's first postulate by APP's first postulate, is that there is not any fundamental principle or law of Nature, which states that the physically meaningful metric must be the WGI product $\mathfrak{g}_{\mu\nu}=\phi^2g_{\mu\nu}$ and not the metric $g_{\mu\nu},$ itself, which is the one that undergoes the Weyl gauge transformations. According to our approach, the different gauges bear different physical and observational consequences. This is unlike the conventional approach, according to which the different gauges must be identified, so that the gauge choice has no physical consequences. 

A statement such like: ``the scalar field can be gauged away'' through an appropriate conformal transformation, although true is misleading. That the conformal transformation \eqref{gauge-t-a0} transforms, for instance, the Lagrangian \eqref{gauge-a-lag} into the Einstein-Hilbert Lagrangian \eqref{eh-lag}, i. e., that \eqref{gauge-t-a0} relates an arbitrary gauge ${\cal G}_a$ with the GR gauge ${\cal G}_0$, is true. But these gauges can not be identified: ${\cal G}_a\neq{\cal G}_0$, i. e., they do not represent equivalent descriptions of the gravitational laws. In other words, although the gauges ${\cal G}_a$ and ${\cal G}_0$ are linked by the conformal transformation \eqref{gauge-t-a0}, meaning that they are linked by a mathematical equivalence relationship, according to the AAP those gauges are not physically equivalent since they represent different, experimentally checkable phenomenology. 

%In item 6, for instance, it is stated that the quantity $\phi\bar{\Psi}\Psi$, is the one with Weyl gauge invariant (physical) meaning. But there is not any principle which obliges WGI quantities to be the physically meaningful ones. In contrast, if we adhere to the minimal coupling principle, this is the density $\bar{\Psi}\Psi$, the one with the physical meaning. The statement that ``A gauge choice cannot have observable consequences, so that, for instance, the redshift measurements are gauge-independent,'' is obviously tied to the conventional approach to Weyl gauge symmetry.

%------------------------item 6----------------------

The critique in sixth item originates from ignorance of the underlying postulates as well. The frame covariant formalism developed in \cite{comment-3} for multifield inflation is based in previous works, among which we may cite the references \cite{comment-5} and \cite{comment-6}. This formalism was primarily intended to deal with the cosmological multifield perturbation theory. The scalar fields are treated as coordinates on a field-space manifold. Fields transformations, including conformal transformation of the metric and fields reparametrizations, are considered as coordinate transformations in the space of fields. However, spacetime coordinates and fields carry different physical significance, so that the similitude can not be pursued much further. In particular, the WGS is not a dynamical symmetry since the associated Noether current identically vanishes \cite{jackiw-2014, oda-2022, rodrigo-arxiv}. The statement that ``the change of frame... cannot change the classical physics'' is itself a postulate which leads to some physical consequences. The opposite postulate leads to different physical consequences. Only experiment can either support or rule out these postulates.

%----------------------item 7------------------------

The comment in seventh item are influenced by the question about the (in)equivalence of the conformal frames in which scalar-tensor theories of gravity can be formulated \cite{dicke-1962, faraoni-2007, fujii-book, quiros-rev, faraoni_rev_1997, faraoni_ijtp_1999, ct-ineq-nojiri, ct-ineq-brisc, sarkar_mpla_2007, ct-ineq-capoz, quiros-grg-2013, fatibene-2014, ct(inequiv)-7, ct-ineq-brooker, ct-ineq-baha, ct(inequiv)-8, paliatha-2, quiros-ijmpd-2020, ct-ineq-2020, faraoni-book, flanagan-2004, sotiriou_etall_ijmpd_2008, saal_cqg_2016, ct-1, ct(fresh-view)-3, indios_consrvd_prd_2018, thermod_prd_2018, shtanov-2022}. However, this issue has nothing to do with the one about the understanding of WGS in gauge invariant gravitational theories. The gravitational laws within the framework of STT are not invariant under the conformal transformations of the metric: these laws do not look the same in the Jordan frame as in the Einstein frame, for instance. The formalism developed in \cite{comment-4} overcomes this difficulty of scalar-tensor theories in their standard formulation, however, in this formalism the coupling functions ${\cal A}={\cal A}(\Phi)$ and ${\cal B}={\cal B}(\Phi)$ in the gravitational action,

\begin{align} S=\int d^4x\sqrt{-g}\left[M^2_\text{pl}{\cal A}R-{\cal B}(\der\Phi)^2-2{\cal V}(\Phi)\right],\label{stt}\end{align} are transformed (see equations(4) and (7) of Ref. \cite{comment-4}). These couplings enter in the expression for the Newton's constant which is measured in Cavendish-type experiments \cite{quiros-rev, comment-7}:

\begin{align} 8\pi G_0=\frac{1}{\Phi_0}\left[\frac{3+2\omega_0+e^{-M_0r}}{3+2\omega_0}\right],\label{newton-c}\end{align} where,

\begin{align} &\omega(\Phi)\equiv\frac{{\cal A}(\Phi){\cal B}(\Phi)}{M^2_\text{pl}{\cal A}^2_{,\Phi}},\nonumber\\
&M^2(\Phi)\equiv\frac{\Phi{\cal V}(\Phi)}{3+2\omega(\Phi)},\nonumber\end{align} while $\omega_0=\omega(\Phi_0)$, $M_0=M(\Phi_0)$ and $\Phi_0$ is the background value of the scalar field $\Phi$. Since neither $\omega(\Phi)$ nor $M^2(\Phi)$ are invariant under the transformations (2)-(7) in \cite{comment-4}, then these transformations link representations where the laws of gravity look the same, but which differ in the measured value of the gravitational constant.\footnote{The purpose of \cite{comment-4} was to express the slow-roll parameters and the relevant observables in terms of frame and reparametrization invariant quantities, in order to be able to compare apparently different models of inflation and not to develop a gauge invariant theory of gravity.} A similar argument can be applied to the formalism developed in \cite{comment-3} (see our reply to comment in sixth item.) 

In the comment in seventh item, there is a reference to \cite{thooft-2015}, in order to support the statement that ``the physics is frame independent.'' However, in that bibliographic reference it is stated that ``...we fail to understand the symmetry of the scale transformations.'' Then, it is pointed out in \cite{thooft-2015} that ``Since the world appears not to be scale invariant, this symmetry, if it exists, must be spontaneously broken... It is this implementation of the symmetry that we should attempt to construct from the evidence we have.'' The point of view that WGS must be a broken symmetry (CAP's second postulate,) is contrary to the one undertaken in the AAP, which is shared by \cite{comment-2, shaukat-1, shaukat-2, shaukat-3}. According to the latter work, ``Mathematically, Weyl invariance seems not to be a manifest symmetry of nature. But neither are $U(1)$ or general coordinate invariance or other symmetries, if one picks a particular gauge. Perhaps physical theories have been written in a special Weyl gauge thus breaking the Weyl symmetry!... theories with dimensionful couplings are the gauge fixed versions of certain Weyl invariant ones.'' 

%-----------------item 8----------------------

The statement in item eighth that ``The ”fifth force” that would arise upon a conformal transformation of timelike geodesics, is an apparent effect.'' is, at least questionable, until statements are made in regard to the geometric structure of spacetime background. According to the CAP, which is the most widespread point of view on WGS, assuming a Riemannian spacetime background $({\cal M}_4,g_{\mu\nu})$, the timelike geodesics, written in terms of the gauge invariant metric $\mathfrak{g}_{\mu\nu}=\phi^2g_{\mu\nu}$, are given by \eqref{triv-tl-geod}:

\begin{align} \frac{d^2x^\alpha}{d\mathfrak{s}^2}+\mathfrak{C}^\alpha_{\mu\nu}\frac{dx^\mu}{d\mathfrak{s}}\frac{dx^\nu}{d\mathfrak{s}}+\mathfrak{h}^{\alpha\mu}\frac{\der_\mu\phi}{\phi}=0,\nonumber\end{align} where the term $\propto\der_\mu\phi/\phi$, spoils the gauge invariance. This term can not be removed through any point-dependent transformation of the affine parameter. This is what we call as the fifth force term, since it deviates the motion of timelike particles from being a geodesic of Riemann spacetime $({\cal M}_4,\mathfrak{g}_{\mu\nu})$. Since Eq. \eqref{triv-tl-geod} is written in terms of WGI quantities, the fifth force term is present in any frame and so, it is not an apparent effect. If assume, on the contrary, a WIG spacetime $({\cal M}_4,g_{\mu\nu},\phi)$, then the timelike fields follow geodesics of $\tilde W_4$ space, which in terms of the WGI metric tensor $\mathfrak{g}_{\mu\nu},$ amount to timelike geodesics of Riemann spacetime $({\cal M}_4,\mathfrak{g}_{\mu\nu})$:

\begin{align} \frac{d^2x^\alpha}{d\mathfrak{s}^2}+\mathfrak{C}^\alpha_{\mu\nu}\frac{dx^\mu}{d\mathfrak{s}}\frac{dx^\nu}{d\mathfrak{s}}=0.\nonumber\end{align} It is seen that no fifth force arises in this case.

%%%%%%%%%%%%%%%%%%%%%%%%%%%%%%%%%%%%%%%%%

\section{Discussion}\label{sect-discu}

%%%%%%%%%%%%%%%%%%%%%%%%%%%%%%%%%%%%%%%%

The mere concept of symmetry is at the core of our discussion in this paper. Is WGS an actual symmetry or it is not? If not, then what do local scale transformations \eqref{gauge-t} mean? It is widely accepted that gauge symmetry is not a symmetry at all but it is an ambiguity in the field definition or, in other words, it is a mathematical redundancy. This assumption is tightly tied to the requirement that only Weyl gauge invariant quantities are physically meaningful. But, is this a principle of Nature, or just a useful postulate? This is one of the underlying postulates of the conventional approach to WGS. One of its consequences is that the physically meaningful metric is the Weyl gauge invariant combination $\mathfrak{g}_{\mu\nu}=\phi^2g_{\mu\nu}.$ However, in terms of this physical metric, local scale transformations \eqref{gauge-t} amount to the unity transformation: $\mathfrak{g}_{\mu\nu}\rightarrow\mathfrak{g}_{\mu\nu},$ i. e., to no transformation at all. Means that the redundancy postulate trivializes the Weyl gauge transformations. 

Our unconventional approach to WGS challenges the redundancy postulate: it is assumed that WGT link different gauges, representing different classical gravitational states which are characterized by different physical properties. We follow a minimal coupling recipe according to which the metric $g_{\mu\nu}$ and related quantities, such as the diffeomorphism invariants of $\tilde W_4$ space, $ds^2=g_{\mu\nu}dx^\mu dx^\nu$, $\hat R=g^{\mu\nu}\hat R^\lambda_{\;\;\mu\lambda\nu}$, $\hat K=\hat R_{\sigma\mu\lambda\nu}\hat R^{\sigma\mu\lambda\nu},$ etc. are physically meaningful quantities. The gauge invariants: $\phi^2g_{\mu\nu},$ , $\phi^{-2}\hat R,$ $\phi^{-4}\hat K$, etc., which have themselves physical meaning, relate quantities in the different gauges. Take, for instance two gauges: the arbitrary gauge ${\cal G}_a$ and the GR gauge ${\cal G}_0$. We have that:

\begin{align} \phi_a^{-2}\hat R_a=M^{-2}_\text{pl}R,\;\phi_a^{-4}\hat K_a=M^{-4}_\text{pl}K_\text{gr},\nonumber\end{align} where $R$ is the curvature scalar in the GR gauge, $K_\text{gr}=R_{\sigma\mu\lambda\nu}R^{\sigma\mu\lambda\nu}$ is the GR's Kretschmann invariant and we have set $\phi_0=M_\text{pl}$. From these equations it follows that $\hat R_a=(\phi_a/M_\text{pl})^2R$ and $\hat K_a=(\phi_a/M_\text{pl})^4K_\text{gr}.$ Hence, for known $R$ and $K_\text{gr}$ in the GR gauge, different choices of the scalar $\phi_a$, lead to the expressions for the curvature scalar and the Kretschmann invariant in different gauges. One can imagine $N$ identical copies of the Universe: ${\cal U}$, and that the different gauges ${\cal G}_a$, $a=0,1,2,...,N$ ($N\rightarrow\infty$,) correspond to the potentially different representations of the Universe: $\mathfrak{U}_a:={\cal G}_a[{\cal U}]$. The collection $\{\mathfrak{U}_0, \mathfrak{U}_1, ...,\mathfrak{U}_N\}$ was called in Ref. \cite{quiros-prd-2023-1} as the ``many-worlds'' interpretation of WGS. Since the different representations $\mathfrak{U}_a$ carry different physical consequences, these can be experimentally differentiated. The role of experiment is to pick out the representation which better describes the existing amount of observational data. The winning representation is the one corresponding to our world.

Does our understanding of gauge symmetry actually correspond to some type of symmetry? A symmetry of the system, for instance a global symmetry, takes one physical state of the system into a different physical state with the same properties. In line with the CAP, gauge symmetry takes one state of the system into itself, because the gauge transformation is, in this case, a redundancy of the mathematical description. Meanwhile, according to the APP, the WGT \eqref{gauge-t} take one state of the system into a different state, with different physical properties. This explains, in particular, why the associated conserved Noether current is identically vanishing \cite{jackiw-2014, oda-2022, rodrigo-arxiv}: there is nothing to be conserved there! Although it is technically incorrect to call the WGS as a symmetry, the fact is that the Lagrangian \eqref{tot-lag}, as well as the derived EOM \eqref{grav-eom}, are invariant under the Weyl gauge transformations. This means that the laws of gravity are the same in every gauge, even if the different gauges represent different states with different physical properties. 

In order to better understand the above state of affairs, let us trace a parallel with Lorentz invariance in special relativity (SR). The laws of electromagnetism, for instance, the inhomogeneous Maxwell equation: $\der_\lambda F^{\mu\lambda}=4\pi j^\mu$, where $F_{\mu\nu}$ is the EM field strength and $j^\mu$ is the current density, are invariant under the Lorentz transformations:

\bea &&F^{\mu\nu}\rightarrow\Lambda^\mu_{\;\;\sigma}\Lambda^\nu_{\;\;\lambda}F^{\sigma\lambda},\;j^\mu\rightarrow\Lambda^\mu_{\;\;\sigma}j^\sigma,\nonumber\\
&&dx^\mu\rightarrow\Lambda^\mu_{\;\;\nu}dx^\nu,\;\eta_{\mu\nu}\rightarrow\Lambda^\sigma_{\;\;\mu}\Lambda^\lambda_{\;\;\nu}\eta_{\sigma\lambda},\label{lorentz-t}\eea where $\eta_{\mu\nu}$ represents the Minkowski metric and $\Lambda^\mu_{\;\;\nu}$ is the Lorentz boost. The latter links two different inertial reference frames (IRF), say: the rest frame ${\cal F}_0$ and an arbitrary IRF ${\cal F}_a$, which is in relative motion with constant speed ${\bf u}_a$, with respect to the rest frame. Invariance of the EM laws under \eqref{lorentz-t} means that these laws are the same in every IRF. Yet, the observers in different IRFs see different descriptions of the same physical phenomenon. It may happen, for instance, that in the rest frame ${\cal F}_0$, the observer, which is equipped with appropriated measuring instruments, measures the electric field with components $E^{(0)}_i=F^{(0)}_{i0}$, exclusively, since the remaining space-space components of the field strength vanish: $F_{(0)}^{ij}=0$. Under \eqref{lorentz-t} the following specific transformations take place: $F_{(a)}^{ij}=\Lambda^i_{\;\;0}\Lambda^j_{\;\;k}F_{(0)}^{0k}$. Hence, provided that $\Lambda^i_{\;\;0}\neq 0$ and $\Lambda^j_{\;\;k}\neq 0$ for some $i,j,k$, the space-space components of the field strength in the IRF ${\cal F}_a$ do not vanish: $F_{(a)}^{ij}\neq 0$. In consequence, an observer in ${\cal F}_a$, would measure not only the electric field, but also the magnetic field with (at least one) nonvanishing component $B^{(a)}_i=\epsilon_{ijk}F_{(a)}^{jk}/2\neq 0$, where $\epsilon_{ijk}$ is the Levi-Civita symbol. The answer to the question, does the magnetic field ${\bf B}$ really exist? is observer dependent. For the observer in the rest IRF, ${\bf B}=0$, so that in ${\cal F}_0$ not any magnetic field can be measured. However, for the observer in the IRF ${\cal F}_a$, a nonvanishing magnetic field: ${\bf B}\neq 0$, can be measured. This magnetic field may be used to trigger other physical phenomena as, for instance, the Zeeman effect \cite{zeeman}. This can be summarized in the following way: Although the EM laws are the same in any IRF, the Lorentz transformations link two different physical states of the same system.

%%%%%%%%%%%%%%%%%%%%%%%%%%%%%%%%

\section{Conclusion}
\label{sect-conclu}

%%%%%%%%%%%%%%%%%%%%%%%%%%%%%%%%

In the present paper we have discussed about a subject that usually passes unnoticed: the role of postulates in the understanding of gauge symmetry. We found that the conventional approach to Weyl gauge symmetry, called here as the CAP, is based on two postulates: 1) WGS is a mathematical redundancy and 2) WGS is broken symmetry in Nature. If these postulates are replaced by other assumptions, an alternative approach arises which shares the same mathematical basis. In this work we have replaced the above postulates by quite the opposite ones: 1) WGS is not a redundancy of the mathematical description of the gravitational laws and 2) WGS is not a broken symmetry. The resulting alternative approach to WGS is called here as the APP. According to this approach the metric itself -- the one which undergoes the conformal transformation in the WGT \eqref{gauge-t} -- has independent physical meaning. This means that the WGT links different gauges which amount to different representations of our world or, equivalently, to different worlds. In consequence, there is not any Noether current that must be conserved during passage from one world to another. Experiments, in particular the redshift measurements, can differentiate between the different gauges/worlds.

%-------------------acknowledgments----------------------

{\bf Acknowledgments.} The author thanks M. Oquendo for useful comments and FORDECYT-PRONACES-CONACYT for support of the present research under grant CF-MG-2558591.  

%-------------------------------------------------------

%%%%%%%%%%%%%

\appendix

%%%%%%%%%%%%

%---------------------------------------------------------------------

%%%%%%%%%%%%%%%%%%%%%%%%%%%%%%%%%%%%%%%%%%%%%%%%%%%

\section{Varying mass and fifth force in $V_4$}
\label{app-a}

%%%%%%%%%%%%%%%%%%%%%%%%%%%%%%%%%%%%%%%%%%%%%%%%%%%

Varying masses in $V_4$ would entail, necessarily, the presence of a fifth force. Actually, when point-dependent masses are considered in $V_4$, then $m=m(x)$ can not be taken out of the action integral: $S=\int mds.$ From this action the following (time-like) EOM in $V_4$ space, can be derived:

\begin{align} \frac{d^2x^\alpha}{ds^2}+\left\{^\alpha_{\mu\nu}\right\}\frac{dx^\mu}{ds}\frac{dx^\nu}{ds}=\frac{\der_\mu m}{m}h^{\mu\alpha},\label{time-geod}\end{align} where the term $h^{\nu\mu}\der_\mu m/m$ accounts for the variation of mass during parallel transport and 

\begin{align} h^{\mu\alpha}:=g^{\mu\alpha}+u^\mu u^\alpha=g^{\mu\alpha}-\frac{dx^\mu}{ds}\frac{dx^\alpha}{ds},\label{orto-proj}\end{align} is the orthogonal projection tensor, which projects any vector or tensor onto the hypersurface orthogonal to the four-velocity vector $u^\mu=dx^\mu/d\tau$. The RHS of \eqref{time-geod}, in particular the term $\propto\der^\mu\ln m$, which can not be eliminated by any affine parameter transformation whatsoever, is to be regarded as a fifth force which deviates the motion of the time-like test particle from being a geodesic in $V_4$.

%%%%%%%%%%%%%%%%%%%%%%%%%%%%%%%%%%%%%%%%%%%%

\section{Autoparallels in $W_4$ space}
\label{app-b}

%%%%%%%%%%%%%%%%%%%%%%%%%%%%%%%%%%%%%%%%%%%%

Here we shall derive the time-like autoparallels of $W_4$ space. Hence, we assume that the following nonmetricity law takes place: $\hat\nabla_\alpha g_{\mu\nu}=-Q_\alpha g_{\mu\nu}$. At the end we shall go to the particular case of interest in this paper, when the vectorial nonmetricity is the gradient of a scalar field: $Q_\mu=\der_\mu\phi^2/\phi^2$. 

In order to make the gauge symmetry compatible with well-known derivation rules and with the inclusion of fields into $W_4$, it is necessary to introduce the Weyl gauge derivative operators in a way that is equivalent to the one appearing in \cite{dirac-1973, utiyama-1973, maeder-1978}. Let ${\bf T}$ be a $(p,q)$-tensor in $W_4$, with coordinate components $T^{\alpha_1\alpha_2\cdots\alpha_p}_{\beta_1\beta_2\cdots\beta_q}$ and with conformal weight $w({\bf T})=w$, so that under \eqref{gauge-t}: ${\bf T}\rightarrow\Omega^w{\bf T}$. Then, the Weyl gauge derivative and Weyl gauge covariant derivative, respectively, are defined as it follows:

\bea &&\der^*_\alpha{\bf T}:=\der_\alpha{\bf T}+\frac{w}{2}Q_\alpha{\bf T},\nonumber\\
&&\hat\nabla^*_\alpha{\bf T}:=\hat\nabla_\alpha{\bf T}+\frac{w}{2}Q_\alpha{\bf T}.\label{gauge-der}\eea The above definitions warrant that the gauge derivative and the gauge covariant derivative, transform like the geometrical object itself, i. e., under \eqref{gauge-t}: $\der^*_\alpha{\bf T}\rightarrow\Omega^w\der^*_\alpha{\bf T},$ $\hat\nabla^*_\alpha{\bf T}\rightarrow\Omega^w\hat\nabla^*_\alpha{\bf T}.$ 

Let ${\cal C}$ be a curve in $W_4$ that is parametrized by the affine parameter $\xi$. I. e., ${\cal C}$ has coordinates $x^\mu(\xi)$. The gauge covariant derivative along the path $x^\mu(\xi)$ is given by the following operator:

\bea \frac{D^*}{d\xi}:=\frac{dx^\mu}{d\xi}\hat\nabla^*_\mu.\label{gcov-der-path}\eea The parallel transport of given tensor ${\bf T}$ with coordinate components $T^{\alpha_1\alpha_2\cdots\alpha_p}_{\beta_1\beta_2\cdots\beta_q}$, along the path $x^\mu(\xi)$, is defined by the following requirement (this definition coincides with the one in \cite{dirac-1973, maeder-1978}):

\bea \frac{D^*{\bf T}}{d\xi}:=\frac{dx^\mu}{d\xi}\hat\nabla^*_\mu{\bf T}=0\;\Leftrightarrow\;\frac{D^*}{d\xi}T^{\alpha_1\alpha_2\cdots\alpha_p}_{\beta_1\beta_2\cdots\beta_q}=0.\label{gcov-der-path-t}\eea 

In Weyl space $W_4$ the ``timelike'' autoparallels are those curves along which the gauge covariant derivative of the tangent four-velocity vector ${\bf u}$, vanishes. Here $u^\mu=dx^\mu/d\tau$ are the coordinate components of ${\bf u}$ and, as long as this does not cause loss of generality, we chose the proper time $\tau$ to be the affine parameter along the autoparallel curve. The conformal weight of the four-velocity vector $w({\bf u})=-1$. In other words, the autoparallel curves satisfy: $D^*{\bf u}/d\tau=u^\mu\hat\nabla^*_\mu{\bf u}=0,$ or, explicitly,
 
\begin{align} \frac{du^\alpha}{d\tau}+\Gamma^\alpha_{\;\;\mu\nu}u^\mu u^\nu-\frac{1}{2}Q_\mu u^\mu u^\alpha=0.\label{u-autop}\end{align} In terms of the arc-length $d\tau\rightarrow ids$ this equation reads:

\bea \frac{d^2x^\alpha}{ds^2}+\{^\alpha_{\mu\nu}\}\frac{dx^\mu}{ds}\frac{dx^\nu}{ds}-\frac{1}{2}Q_\mu h^{\mu\alpha}=0,\label{time-auto-p'}\eea where $h_{\mu\nu}$ is defined in \eqref{orto-proj}. If in \eqref{time-auto-p'} we make the substitution: $Q_\mu=\der_\mu\phi^2/\phi^2=2\der_\mu\phi/\phi$, then equation \eqref{time-auto-p} is obtained.

%%%%%%%%%%%%%%%%%%%%%%%%%%%%%%%%%%%%%%%%%%%%%%%%%

\section{Mass variation in $W_4$ space}
\label{app-c}

%%%%%%%%%%%%%%%%%%%%%%%%%%%%%%%%%%%%%%%%%%%%%%%%%

Let us consider the four-momentum vector ${\bf p}=m{\bf u}$ of a point particle, with components: $p^\mu=mu^\mu$, where $u^\mu=dx^\mu/d\tau$ is the particle's four-velocity vector and its mass is a function of the spacetime coordinates $m=m(x)$. Let us further consider the parallel transport of the four-momentum ${\bf p}$, along the path $x^\mu(\tau)$:

\begin{align} \frac{D^*{\bf p}}{d\tau}=\frac{D^*m}{d\tau}{\bf u}+m\frac{D^*{\bf u}}{d\tau}=0.\label{app-b-1}\end{align} Taking into account the autoparallel's equation \eqref{u-autop} and the definition of the gauge covariant derivative along given path $x^\mu(\tau)$ (see equation \eqref{gcov-der-path},) we get that,

\begin{align} \frac{D^*m}{d\tau}=\frac{dx^\mu}{d\tau}\der^*_\mu m=\frac{dx^\mu}{d\tau}\left(\der_\mu m-\frac{m}{2}Q_\mu\right)=0,\label{app-b-2}\end{align} from where the following equation is obtained:

\begin{align} \frac{\der_\mu m}{m}=\frac{1}{2}Q_\mu.\label{app-b-3}\end{align} If replace $Q_\mu\rightarrow\der_\mu\phi^2/\phi^2$ in this equation, one obtains equation \eqref{mass-autop} and, consequently, equation \eqref{mass-phi-rel}. In general, in Weyl space where vectorial nonmetricity takes place, from \eqref{app-b-3} the following relationship arises:

\begin{align} m(x,{\cal C})=m_0 e^{\frac{1}{2}\int_{\cal C} Q_\mu dx^\mu},\label{app-b-4}\end{align} where $m_0$ is an integration constant and ${\cal C}$ is the path of parallel transport. In this case the mass of a point-particle, being the length of the four-momentum, varies along the path of parallel transport. Means that it depends not only on the spacetime point but, also, on the followed path $m(x,{\cal C})$. In $\tilde W_4$ space, where $Q_\mu=2\der_\mu\phi/\phi$, the mass depends only on the spacetime point $m=m(x)$, instead.

%%%%%%%%%%%%%%%%%%%%%%%%%%%%%

%%%%%%%%%%%%%%%%%%%%%%%%%%

%%%%%%%%%%%%%%%%%
%%%%%%%%%%%%%%%%%

\end{document}